\documentclass[12pt]{amsart}

\usepackage{amssymb}
\usepackage{amsmath,amsthm,amsfonts}  
 \usepackage{graphicx}
\usepackage{amssymb}
\usepackage{pdfsync}
\usepackage{fancyhdr}
\usepackage[applemac]{inputenc}
\usepackage{mathabx}
\usepackage{mathtools}
\usepackage{subfig}
\usepackage{bbm}
\usepackage{float}
\usepackage[T1]{fontenc}

\usepackage{color}
\usepackage{xcolor}
\usepackage[breaklinks,colorlinks,backref]{hyperref}
\hypersetup{
    colorlinks, 
    linktoc=all, 
    linkcolor=red,  
  citecolor=hyptxt,
  urlcolor=blue}
  \hypersetup{
  citebordercolor=,
  filebordercolor=green,
  linkbordercolor=blue
}
\definecolor{hyptxt}{rgb}{0.7, 0.4, 0.9}
\usepackage{bibtopic}

\newcommand\symbolwithin[2]{%
  {\mathmakebox[\widthof{\ensuremath{{}#2{}}}][c]{{#1}}}}
\newcommand{\sI}{\mathbbm{1}}

\definecolor{hervecolor}{rgb}{0.8,0,0.7}
     
\newcommand{\ket}[1]{|\kern.3ex#1\kern.3ex\rangle}
\newcommand{\bra}[1]{\langle\kern.3ex #1 \kern.3ex|}
\newcommand{\scalar}[2]{\langle\kern.3ex #1 \kern.3ex|\kern.3ex#2\kern.3ex\rangle}

\newcommand{\ii}{\mathsf{i}}

\def\R{\mathbb{R}}

\def\C{\mathbb{C}}

\def\lg{\langle }
\def\rg{\rangle }

\def\ud{\mathrm{d}}
\def\sfH{\mathsf{H}}

\def\sfP{\mathsf{P}}

\def\mFs{\mathfrak{F_s}}

\def\omFs{\overline{\mFs}}
\def\mfs{\mathfrak{f_s}}
\def\omfs{\overline{\mathfrak{f_s}}}

\numberwithin{equation}{section}

\begin{document}
\date{\today}
 
\title[Weyl-Heisenberg quantization \& Variable mass]{Quantum and semi-classical aspects of confined systems with variable mass}
\author[Gazeau-Hussin-Moran-Zelaya]{
Jean-Pierre Gazeau$^{\mathrm{a}}$, V\'eronique Hussin$^{\mathrm{b},\mathrm{c}}$, James Moran$^{\mathrm{b},\mathrm{d}}$, and Kevin Zelaya$^{\mathrm{b}}$}

\address{\emph{$^{\mathrm{a}}$ Universit\'e de Paris, CNRS}\\ \emph{Astroparticule et Cosmologie} \emph{F-75006 Paris, France}}


\address{\emph{$^{\mathrm{b}}$ Centre de Recherches Math\'ematiques}\\ \emph{Universit\'e de Montr\'eal, Montr\'eal QC H3C 3J7, Canada}}

\address{\emph{$^{\mathrm{c}}$ D\'epartement de Math\'ematiques et de Statistique}\\ \emph{Universit\'e de Montr\'eal, Montr\'eal QC H3C 3J7, Canada}}

\address{\emph{$^{\mathrm{d}}$ D\'epartement de Physique}\\ \emph{Universit\'e de Montr\'eal, Montr\'eal QC H3C 3J7, Canada}}

\email{e-mail:
gazeau@apc.in2p3.fr, veronique.hussin@umontreal.ca, james.moran@umontreal.ca, zelayame@crm.umontreal.ca}

{\abstract{We explore the quantization of classical models with position-dependent mass (PDM) terms constrained to a bounded interval in the canonical position. This is achieved through the Weyl-Heisenberg covariant integral quantization by properly choosing a regularizing function $\Pi(q,p)$ on the phase space that smooths the discontinuities present in the classical model. We thus obtain well-defined operators without requiring the construction of self-adjoint extensions. Simultaneously, the quantization mechanism leads naturally to a semi-classical system, that is, a classical-like model with a well-defined Hamiltonian structure in which the effects of the Planck's constant are not negligible. Interestingly, for a non-separable function $\Pi(q,p)$, a purely quantum minimal-coupling term arises in the form of a vector potential for both the quantum and semi-classical models.}}

\maketitle

\tableofcontents

\section{Introduction}
\label{intro}


The aim of this work is to study the extent to which the mass of a system depends on the quantum regime on geometric confinement in the configuration space.    The procedure is illustrated with one of the most elementary examples in mechanics, namely the motion of a particle in a subset $E$ of the  line $\R$.
 The quantum versions of this classical model are established by following  the so-called covariant Weyl-Heisenberg (WH)integral quantization \cite{bergaz14,gazeau18,becugaro17}, a procedure which is built from a function $\Pi(q,p)$ whose the symplectic Fourier transform
\begin{equation*}
 \mathfrak{F_s}[\Pi](q,p):= \int_{\R^2}e^{-\frac{\ii}{\hbar}(qp^{\prime}-pq^{\prime})}\, \Pi(q^{\prime},p^{\prime})\,\frac{\ud q^{\prime}\,\ud p^{\prime}}{2\pi \hbar}  
\end{equation*}
 is assumed to be a distribution probability on the plane viewed as the phase space for the motion of a particle on the line.  
   
In a nutshell,  the canonical quantization of classical observables $f(q,p)$ consists in the replacements 
\begin{align}
\R^2\ni (q,p)  &\mapsto \ \mbox{self-adjoint}\ (Q,P)\,, \\
f(q,p) &\mapsto f(Q,P) \mapsto (\mathrm{Sym}f)(Q,P)\, ,
\end{align}
where $\mathrm{Sym}$ stands for a certain choice of symmetrisation  of the operator-valued function. One should notice that the \textit{canonical commutation relation} (CCR)
\begin{equation}
\label{ccr}
[Q,P] \equiv QP-PQ= \ii \hbar \sI\, ,
\end{equation}
holds true with (essentially) self-adjoint $Q$, $P$,  only if both have continuous spectrum $(-\infty,+\infty)$.  

Clearly, this  scheme does not encompass  the cases of  other phase space geometries, like  barriers or other  impassable boundaries. Moreover, the quantization of a Hamiltonian with variable mass, $H(q,p) = p^2/2m(q) + \cdots$ raises the well known ordering problem, see for instance \cite{jmll92,jmll95,gazkoi19} and references therein. 
 
This concept of variable mass is a natural outcome of the so-called \textit{shadow} Galilean invariance \cite{jmll74}. For the motion of an interacting massive  particle on the line, it is reasonable to impose that  no discrimination is possible instantaneously between a free and an interacting system.  
After introducing the evolution parameter $t$,  one shows that shadow Galilean   dynamics is  ruled by  Hamiltonians of the general form,  
\begin{equation}
\label{hamiltGal1}
\begin{split}
\sfH^{\mathrm{gen}}(q,p;t) &= \frac{1}{2m(q)}(p - A(q;t))^2 + U(q;t)\\
&=  \frac{p^2}{2m(q)} -\frac{p}{m(q)}\,A(q;t) + \frac{A^2(q;t)}{2m(q)} + U(q;t) \\ &\equiv L_2(q;t)\,p^2 + L_1(q;t)\,p + L_0(q;t)\, , 
\end{split}
\end{equation}
to which the WH integral quantization applies easily, and plays in general a regularizing role, depending on the choice of the function $\Pi(q,p)$. Note that this choice will dispel the ordering ambiguity due to the presence of the variable mass $m(q)$.  Of course, changing $\Pi(q,p)$ yields in general another ordering. 

Now, imposing to a particle with a constant mass $m_0$ to move inside a subset $E\subset \R$ entails that its Hamiltonian has to be multiplied by the characteristic function 
\begin{eqnarray}
\chi_{E} (q)= 
\left\{
\begin{array}{cl}
1 & q\in E \\
0 & {\rm otherwise}
\end{array} \, .
\right.
\end{eqnarray}
Hence, the Hamiltonian becomes the truncated observable
\begin{equation}
\label{Hchi}
H(q,p) = \frac{p^2}{2m_0} + \cdots \quad\mapsto \quad \chi_{E} (q)H(q,p)= \frac{p^2}{2m(q)} + \cdots\, ,\,.
\end{equation}
where 
\begin{equation}
\label{momq}
 \quad m(q)= \frac{m_0}{\chi_{E} (q)}\,.
\end{equation}
Clearly, this variable mass $m(q)$ is a singular function, since it becomes infinite  at the boundary of $E$. Generalising the procedure introduced  in the previous work \cite{gazkoi19}, we show in this paper that our quantization method based on the choice of a, at least, continuous function $\Pi (q,p)$ may  regularise this singular mass and yield a soft semi-classical Hamiltonian mechanics, say \textit{\`a la Klauder} \cite{klauder12,klauder15}, allowing the particle to escape its confining geometry $E$, at the order $\hbar$.

Covariant Weyl-Heisenberg quantization is presented in Section \ref{CIQWH}.  The quantum counterparts issued from the procedure are made explicit in Section \ref{permissu} for the most common observables.   In Section \ref{weight} we examine the outcomes yielded by the choice of some functions $\Pi(q,p)$. The semi-classical side of the quantization and its probabilistic aspects are described in  Section \ref{probcont}. In Section \ref{klauder} we adapt and generalize   the Klauder's formalism of enhanced quantization through the use of the function $\Pi(q,p)$. The general formalism of the quantization of truncated observables and subsequent semi-classical portraits are developed in Section \ref{qutrunc}. The elementary example of the motion in an interval is comprehensively treated  in Section \ref{interval}. Applications to  problems involving specific potentials are considered in Section \ref{potentials}. In Section \ref{conclu} some ideas of future works are sketched. Appendix \ref{fourier} is devoted to some useful properties of the symplectic Fourier transform.

\section{Covariant integral quantization of the motion on the  line}
\label{CIQWH}

In this section we remind the content of Reference \cite{gazeau18} in which was described the covariant Weyl-Heisenberg integral quantization  of the motion on the line. Precisely, we transform a function $f(q,p)$ into an operator $A_f$ in some separable Hilbert space $\mathcal{H}$ through a linear map which sends the function $f=1$ to the identity operator in $\mathcal{H}$ and which respects the basic translational symmetry of the phase space. A probabilistic content is one of the most appealing outcomes of the procedure.

\subsection{The quantization map}
 
We define the integral quantization of the motion on the  line  as  the linear map
\begin{equation}
\label{qmap}
f(q,p) \mapsto A_f = \int_{\R^2} f(q,p)\, \mathfrak{Q}(q,p)\, \frac{\ud q\, \ud p}{2\pi \hbar}\, . 
\end{equation}
where $\mathfrak{Q}(q,p)$ is a family of operators which solves the identity in $\mathcal{H}$ with respect to the  measure $\ud q\,\ud p/(2\pi \hbar)$,
\begin{equation}
\label{resid}
\int_{\R^2}  \mathfrak{Q}(q,p)\, \frac{\ud q\,\ud p}{2\pi \hbar} = \sI\,. 
\end{equation}
Hence  the identity  $\sI$  is the quantized version of the function $f=1$. 

In addition to \eqref{resid}, we impose  the family $ \mathfrak{Q}(q,p)$ to obey a symmetry condition issued from the homogeneity of the phase space. 
Indeed, the choice of the origin in $\R^2$ is arbitrary. Hence we must have translational covariance in the sense that the quantization of the translated of $f$ is unitarily equivalent to the quantization of $f$ 
\begin{equation}
\label{covtrans1}
U(q_0,p_0)\,A_f \,U(q_0,p_0)^{\dag}= A_{\mathcal{T}(q_0,p_0)f}\, , 
\end{equation}
where
\begin{equation}
\label{troper}
\left(\mathcal{T}(q_0,p_0)f\right)(q,p):= f\left(q-q_0,p-p_0\right)\,.
\end{equation}
So $(q,p)\mapsto U(q,p)$ has to be a unitary, possibly projective,  representation of the abelian group $\R^2$.

Then, from \eqref{covtrans1} and the translational invariance of $\ud q\,\ud p$, the operator valued function  $\mathfrak{Q}(q,p)$ has to obey 
\begin{equation}
\label{covQ}
U(q_0,p_0)\,\mathfrak{Q}(q,p) \,U^{\dag}(q_0,p_0)= \mathfrak{Q}\left(q+ q_0,p+p_0\right) \,.
\end{equation}
A solution to \eqref{covQ} is  found by picking an operator $\mathfrak{Q}_0\equiv \mathfrak{Q}(0,0)$ and write 
\begin{equation}
\label{solQ0}
\mathfrak{Q}\left(q,p\right) := U(q,p)\,\mathfrak{Q}_0 \,U^{\dag}(q,p) \,.
\end{equation}
Then the resolution of the identity holds from Schur's Lemma  \cite{barracz77} if $U$ is irreducible, and if the operator-valued integral  \eqref{resid} makes sense, i.e., if the choice of the fixed operator  $\mathfrak{Q}_0$ is valid.

\subsection{From $\R^2$ to the Weyl-Heisenberg group and its UIR}
  \label{R2WH} 
  In order to find the non-trivial unitary operator $U(q,p)$ involved in \eqref{covtrans1}, we  deal with the Weyl-Heisenberg (WH) group,
\begin{equation}
\label{WHgroup}
\begin{split}
\mathrm{WH} &= \{(s,q,p)\, , \, s\in \R\, , \, (q,p)\in \R^2\}\, , \\
 (s,q,p)(s^{\prime},q,p) &= \left(s+s^{\prime} - \frac{1}{2\hbar}(qp^{\prime}-pq^{\prime}), q+q^{\prime},p+p^{\prime}\right)\, ,
\end{split}
\end{equation}
instead of just the group $\R^2$. 

From von Neumann \cite{vneumann31,perel86,combescure12}, WH has a unique non trivial \textit{unitary irreducible representation} (UIR), up to equivalence corresponding precisely to the arbitrariness  of a constant $k\neq 0$, that we put equal to $\hbar$:
\begin{equation}
\label{UIRWH}
(s,q,p) \mapsto \mathcal{U}(s,q,p)= e^{\ii s}\,U(q,p) \, ,
\end{equation}
where
\begin{equation}
\label{displU}
 U(q,p)= e^{\frac{\ii}{\hbar} (pQ-qP)}\,.
\end{equation}
is the \textit{displacement operator}, and 
where $Q$ and $P$ are the two above-mentionned self-adjoint operators in $\mathcal{H}$ such that $[Q,P]= \ii\, \hbar \sI$.

 \subsection{WH covariant integral quantization(s)}

From Schur's Lemma applied to the WH UIR $\mathcal{U}$, or equivalently to $U$ since $e^{\ii s}$ is just a phase factor, we confirm the  resolution of the identity 
\begin{equation}
\label{resunit2}
\int_{\R^2}  \mathfrak{Q}(q,p)\, \frac{\ud q\, \ud p}{2\pi \hbar} = \sI\,, \quad \mathfrak{Q}(q,p)= U(q,p)\mathfrak{Q}_0 U^{\dag}(q,p)\, , 
\end{equation}
where $\mathfrak{Q}_0$ is the fixed operator introduced in \eqref{solQ0}, whose choice is left to us. Let us prove  that this is  possible if $\mathfrak{Q}_0$ is trace class with unit trace i.e., $\mathrm{Tr}(\mathfrak{Q}_0)=1$. Indeed, let us introduce the function
\begin{equation}
\label{WHtr}
\Pi(q,p) = \mathrm{Tr}\left(U(-q,-p)\mathfrak{Q}_0 \right)\, . 
\end{equation}
This is interpreted as the \textit{Weyl-Heisenberg transform} of operator $\mathfrak{Q}_0$. 

The \textit{inverse WH-transform} exists due to two remarkable properties \cite{bergaz14,becugaro17} of the displacement operator $U(q,p)$,
\begin{equation}
\label{IWHtr}
 \int_{\R^2} U(q,p) \,\frac{\ud q\,\ud p}{2\pi \hbar} = 2{\sf P}\ \mbox{and}\ \mathrm{Tr}\left(U(q,p)\right)= 2\pi \hbar\delta(q,p)\,,\end{equation}
with $\delta(q,p)\equiv\delta(q)\delta(p)$ the two-dimensional Dirac delta distribution~\cite{abraste72,Olv10}, and ${\sf P}= {\sf P}^{-1}$ the parity operator defined as ${\sf P}U(q,p){\sf P}= U(-q,-p)$, and the trace of ${\sf P}$ is put equal to $1/2$. 

One derives from \eqref{IWHtr} the inverse WH-transform of $\Pi(q,p)$:
 \begin{equation}
\label{PiQ0}
  \mathfrak{Q}_0 = \int_{\R^2} U(q,p) \, \Pi(q,p)\,\frac{\ud q\,\ud p}{2\pi \hbar}\, , 
\end{equation}
The function $\Pi(q,p)$ is like a weight, not necessarily normalisable, or even positive. It can be viewed as an apodization \cite{apodi}  on the plane, or a kind of coarse graining  of the phase space.

Equipped with one choice of a traceclass $\mathfrak{Q}_0$,  we can now proceed with the corresponding WH covariant integral quantization map
\begin{equation}
\label{fAf}
 f(q,p) \mapsto A_f \left(\equiv A^{\mathfrak{Q}_0}_f\right) = \int_{\R^2} f(q,p) \mathfrak{Q}(q,p)\, \frac{\ud q \ud p}{2\pi \hbar}\, . 
\end{equation}
In this context, the operator $\mathfrak{Q}_0$ is the quantum version (up to a constant) of the origin of the phase space, identified with the $2\pi\times$ Dirac distribution at the origin. 
\begin{equation}
\begin{split}
\label{quantdirac}
& 2\pi \hbar\delta(q,p) \mapsto A_{\delta}= \mathfrak{Q}_0\, , \\
& 2\pi \hbar\delta(q-q_0,p-p_0) \mapsto A_{\delta_{(q_0,p_0)}}= \mathfrak{Q}(q_0,p_0) \, .
\end{split}
\end{equation}

The probabilistic content of our quantization procedure is better captured if one uses an  alternative quantization formula through the  symplectic Fourier transform, 
  \begin{equation}
\label{symFourqp}
 \mathfrak{F_s}[f](q,p)= \int_{\R^2}e^{-\frac{\ii}{\hbar} (qp^{\prime}-pq^{\prime})}\, f(q^{\prime},p^{\prime})\,\frac{\ud q^{\prime}\,\ud  p^{\prime}}{2\pi\hbar} \, . 
\end{equation}
 It is involutive, $\mathfrak{F_s}\left[\mathfrak{F_s}[f]\right]=  f$ like its \textit{dual} defined 
 as $\overline{\mathfrak{F_s}}[f](q,p)= \mathfrak{F_s}[f](-q,-p)$. See Appendix \ref{fourier} for details. 
 
The equivalent form of  the WH integral quantization \eqref{fAf} reads as
\begin{equation}
\label{quantPi1}
f\mapsto A_f= \int_{\R^2}  U(q,p)\,  \overline{\mFs}[f](q,p)\, \Pi(q,p) \,\frac{\ud q\,\ud p}{2\pi\hbar}\, .  
\end{equation}

From \eqref{fAf} or \eqref{quantPi1} is derived  the action of $A_f$ as the  integral operator  
\begin{equation}
\label{Afint}
L^2(\R,\ud x) \ni \phi(x) \mapsto (A_f \phi)(x) =\int_{-\infty}^{+\infty}\mathrm{d} x^{\prime}\, \mathcal{A}_f(x,x^{\prime})\, \phi(x^{\prime})\, , 
\end{equation}
with  kernel given by
\begin{equation}
\label{AfintK}
 \mathcal{A}_f(x,x^{\prime})= \frac{1}{2\pi\hbar}\int_{-\infty}^{+\infty}\mathrm{d} q\, \widehat{f}_{p}(q,x^{\prime}-x)\, \widehat{\Pi}_{p}\left(x-x^{\prime},q- \frac{x+x^{\prime}}{2}\right)\,.
\end{equation}
Here the symbol $\widehat{f}_{p}$ stands for partial Fourier transform of $f$ with respect its second variable $p$:
\begin{equation}
\label{pFTp}
\widehat{f}_{p}(q,y)= \frac{1}{\sqrt{2\pi\hbar}}\int_{-\infty}^{+\infty}\ud p \, f(q,p)\, e^{-\ii \frac{yp}{\hbar}}\,.
\end{equation}

 \section{Permanent outcomes of WH covariant integral quantizations}  
 \label{permissu}
 
 By permanent outcomes we mean that some basic rules managing the quantum model have a kind of universality, almost whatever the choice of admissible  $\mathfrak{Q}_0$, or its corresponding  apodization $\Pi(q,p)$.
 
 \subsubsection*{Symmetric operators and self-adjointness}
First, we have  the general important outcome:  if $\mathfrak{Q}_0$ is symmetric, i.e. $[\Pi(-q,-p)]^{*}=\Pi(q,p)$ with $z^{*}$ the complex-conjugate of $z$,  a real function $f(q,p)$ is mapped to a symmetric operator $A_f$. Moreover, if  $\mathfrak{Q}_0$ is a positive operator, then  a real semi-bounded  function $f(q,p)$ is mapped to a self-adjoint operator $A_f$ through the Friedrich extension \cite{akhglaz81} of its associated semi-bounded quadratic form.   

\subsubsection*{Position and Momentum}
Canonical commutation rule is preserved:
\begin{equation}
\label{qandp}
A_q = Q + c_0\, , \quad A_p= P+d_0\,, \   \Rightarrow \ \left[A_q,A_p\right]= \ii \hbar \sI\, . 
\end{equation}
This result is actually the direct consequence of the underlying Weyl-Heisenberg covariance when one expresses~\eqref{covtrans1} on the level of infinitesimal generators.

 \subsubsection*{Kinetic energy}
\begin{equation}
\label{p2}
 A_{p^2}= P^2 + e_1\,P + e_0\, . 
\end{equation}
\subsubsection*{Dilation} 
\begin{equation}
\label{qp}
A_{qp} = A_q\,A_p + f_0\, .
\end{equation}
In the above formulas,  the constants $c_0,d_0,e_0,e_1,$ can be easily removed by imposing mild constraints on $\Pi(q,p)$. Moreover, constant $f_0$ can be fixed to $-\ii/2$ in order to get the symmetric dilation operator $(QP + PQ)/2$. 
\subsubsection*{Potential energy} A potential energy $V(q)$ becomes a multiplication operator in position representation.
\begin{equation}
\label{Vq}
A_{V} = \mathfrak{V}(Q)\, , \quad \mathfrak{V}(Q)= \frac{1}{\sqrt{2\pi\hbar}}\,V\ast \overline{\mathcal{F}}[\Pi(0,\cdot)](Q)\, ,
\end{equation}
where $ \overline{\mathcal{F}}$ is the inverse 1-D Fourier transform
\begin{equation}
\label{1dFour}
\mathcal{F}[f](q)=\frac{1}{\sqrt{2\pi\hbar}}\int_{-\infty}^{+\infty}f(p)\,e^{-\frac{\ii}{\hbar} qp}\,\ud p\,, \quad \overline{\mathcal{F}}[f](q)=\mathcal{F}[f](-q)\,,
\end{equation}
 and ``$\ast$'' stands for convolution, both  with respect to the second variable. 

\subsubsection*{Functions of $p$}
If $F(q,p)\equiv v(p)$ is a function of $p$ only, then $A_v$ depends on $P$ only through the convolution:
\begin{equation}
\label{hp}
A_v= \frac{1}{\sqrt{2\pi\hbar}}\,v\ast \mathcal{F}[\Pi(\cdot,0)](P)\, ,
\end{equation}
where the 1-D Fourier transform $ \mathcal{F}$ and the convolution hold  with respect to the first variable.

 \section{Examples of $\Pi$ functions}
 \label{weight}

 The simplest choice is   $\Pi(q,p) = 1$, of course.  Then $\mathfrak{Q}_0 = 2 \sfP$. This no filtering choice yields the popular Weyl-Wigner integral quantization (see \cite{combescure12} and references therein), equivalent to the standard ($\sim$ canonical) quantization. No  regularisation of space or momentum singularity present in the classical model is possible since 
\begin{equation}
\label{VqVQ}
V(q) \mapsto A_V = V(Q)\, , \quad v(p) \mapsto A_v = v(P)\, . 
\end{equation}
This quantization yields the so-called Weyl  ordering \cite{becugaro17}.

Another choice is the Born-Jordan function,  $\Pi(q,p) = \dfrac{\hbar\sin (qp/\hbar)}{qp}$, which presents appealing aspects \cite{cordero_etal15,gosson16}. With this choice Eqs. \eqref{VqVQ}  also hold true.

 If $\mathfrak{Q}_0= |\psi\rg\lg\psi|$, with $\Vert \psi\Vert=1$, then 
 \begin{equation}
\label{psiPiqp}
 \Pi(q,p)= e^{-\ii \frac{qp}{2\hbar}}\left(\mathcal{F}[\psi]\ast\mathcal{F}[\mathrm{t}_{-q}\psi]\right)(p)\, , 
\end{equation}
   where $\mathcal{F}[\mathrm{t}_{-q}\psi](p)= e^{\ii \frac{qp}{\hbar}}\mathcal{F}[\psi](p)$ for $\psi(x)\in L^2(\R,\ud x)$. The corresponding integral kernel is given by
\begin{equation}
\label{Afpsi}
 \mathcal{A}_f(x,x^{\prime})= \frac{1}{\sqrt{2\pi \hbar}}\int_{-\infty}^{+\infty}\mathrm{d} q\, \widehat{f}_{p}(q,x^{\prime}-x)\, \psi(x-q)\,[\psi(x^{\prime}-q)]^{*}
\end{equation}
For instance, let us consider the squeezed vacuum state 
\begin{equation}
\label{sqvacuum}
\vert\xi\rangle=S(\xi)\vert \psi_{\ell} \rangle \,, \quad S(\xi)=e^{-\frac{\xi}{2}a^{\dagger 2}+\frac{\xi^{*}}{2}a^{2}} \,, \xi\in\mathbb{C} \,,
\end{equation}
where $S(\xi)$ stands for the \textit{squeezing operator}~\cite{gerry05}, $\vert\psi_{\ell}\rangle$ the harmonic oscillator ground state
\begin{equation}
\label{grstho}
\lg x|\psi_\ell\rg = \psi_{\ell}(x)= \left( \frac{1}{\pi \ell^2} \right)^{1/4} e^{-\frac{x^2}{2\ell^2}}\,,  
\end{equation}
and $\{a,a^{\dagger}\}$ denote the set of boson operators. The squeezed vacuum state is given, in coordinate representation, by the normalized complex-valued Gaussian function
\begin{equation}
\label{grstho}
\psi_{\ell,\eta}(x)\equiv\langle x \vert \xi\rangle =  \left( \frac{1-\vert \eta \vert^{2}}{\pi \ell^2(1-\eta)^{2}} \right)^{1/4} e^{-\frac{x^2}{2\ell^2}\left(\frac{1+\eta}{1-\eta}\right)} \, , \quad \eta=\frac{\xi}{\vert\xi\vert}\tanh \vert\xi\vert \, ,
\end{equation}
where $\vert \eta \vert<1$, and $\ell$ is a constant that has dimension of position. Its Fourier transform is
\begin{equation*}
\mathcal{F}[\psi_{\ell,\eta}](p)=\left( \frac{1-\vert\eta\vert^{2}}{\pi \wp^2(1+\eta)^{2}}\right)^{1/4}e^{-\frac{p^2}{2\wp^2}\left(\frac{1-\eta}{1+\eta}\right)}\,,
\end{equation*}
where $\wp =\hbar/\ell$ has dimension of momentum. Its associate function $\Pi(q,p)$ is given through \eqref{psiPiqp} by
\begin{equation}
\label{sqgaussPil}
\begin{split}
&\Pi_{\ell,\eta}(q,p)= e^{-\frac{q^{2}}{4\ell^{2}}\left(\kappa_{R}+\frac{\kappa_{I}^{2}}{\kappa_{R}}\right)}e^{-\frac{p^{2}}{4\wp^{2}}\frac{1}{\kappa_{R}}}e^{-\frac{\kappa_{I}}{\kappa_{R}}\frac{qp}{2\hbar}}\\
&\kappa=\frac{1+\eta}{1-\eta} \, , \quad \kappa_{R}=\frac{\kappa+\kappa^{*}}{2} \, , \quad \kappa_{I}=\frac{\kappa-\kappa^{*}}{2i} \,.
\end{split}
\end{equation}
It is worth noting that $\xi=\eta=0$ leads to the Gaussian ground state $\vert \psi_{\ell} \rangle$, and consequently to a function $\Pi(q,p)$ of the form
\begin{equation}
\label{gaussPil}
\Pi_\ell(q,p)\equiv\Pi_{\ell,\eta=0}=e^{-\frac{q^2}{4\ell^2}}e^{-\frac{p^2}{4\wp^2}}\,.
\end{equation}
Its symplectic Fourier transform reads
\begin{equation}
\label{FsgaussPil}
\mathfrak{F_s}\left[\Pi_{\ell}\right](q,p)= 2e^{-\frac{q^2}{\ell^2}-\frac{p^2}{\wp^2}} \,, \quad \ell \wp=\hbar\,.
\end{equation}
$\Pi_\ell(q,p)$ is precisely the function yielding the coherent state or Berezin or anti-Wick quantization, where two parameters are present, $\ell$ and $\hbar$, or equivalently $\wp$ and $\hbar$. The corresponding operator $\mathfrak{Q}_0$ is the orthogonal projector $|\psi_\ell\rg\lg\psi_\ell|$ on the harmonic oscillator ground state.

An easily manageable generalisation of \eqref{gaussPil} concerns  separable functions
\begin{equation}
\label{sepPi}
\Pi(q,p)= \lambda(q)\, \mu(p)\,,
\end{equation} 
where $\lambda$ and $\mu$ are preferably regular, e.g., rapidly decreasing smooth functions. Note its symplectic Fourier transform:
\begin{equation}
\label{sFepPi}
\mathfrak{F_s}[\Pi](q,p)= \mathcal{F}[\mu](q)\,\overline{\mathcal{F}}[\lambda](p)\,.
\end{equation} 
Such an option is  suitable for physical Hamiltonians which are sums of terms like $L(q)\,p^n$, where it  allows  regularisations through convolutions if functions $\lambda$ and $\mu$ are regular enough. 
\begin{equation}
\label{sepweight}
\begin{split}
A_{L(q)\,p^n}&=  \sum_{\substack{
         r,s,t\\
         r+s+t=n}}  2^{-s}\, \binom{n}{r\,s\,t}\,(\ii\hbar)^r\,\lambda^{(r)}(0)\times\\
         &\times(-\ii\hbar)^s \frac{1}{\sqrt{2\pi\hbar}}\,\left(\overline{\mathcal{F}}[\mu]\ast L \right)^{(s)}(Q)\, P^t\, .
\end{split} 
\end{equation} 
Note that $\left(\overline{\mathcal{F}}[\mu]\ast L \right)^{(s)}= \left(\overline{\mathcal{F}}[\mu]\right)^{(s)}\ast L = \overline{\mathcal{F}}[\mu]\ast L^{(s)}$, relations whose  validity depends on the derivability of the factors. 

For the cases $n=0$, $n=1$ and $n=2$, i.e. the most relevant to  Galilean physics, we have, with $T(x):= \frac{1}{\sqrt{2\pi\hbar}}\,\left(\overline{\mathcal{F}}[\mu]\ast L \right)(x)$, 
\begin{equation}
\label{Lpm0}
A_{L(q)} = \lambda(0)\, T(Q)
\, ,
\end{equation}
\begin{equation}
\label{Lpm1}
\begin{split}
A_{L(q)\,p} &= \lambda(0)\, T(Q)\, P + \ii \hbar\,\lambda^{\prime}(0)\, T(Q) -\frac{\ii\hbar}{2}\,\lambda(0)\, T^{\prime}(Q)\\
&= \lambda(0)\, \frac{T(Q)\, P + P\,T(Q)}{2} + \ii\hbar\, \lambda^{\prime}(0)\, T(Q)\, ,
\end{split}
\end{equation}
\begin{align}
\label{Lpm2a} A_{L(q)\,p^2} &= \lambda(0)\, T(Q)\, P^2 + \ii\hbar \,(2\lambda^{\prime}(0)\, T(Q)-  \lambda(0)\, T^{\prime}(Q))\,P \\ 
 \nonumber &+\hbar^2\left(-\lambda^{\prime\prime}(0)\, T(Q)+ \lambda^{\prime}(0)\, T^{\prime}(Q)-\frac{\lambda(0)}{2}\, T^{\prime\prime}(Q)\right)\\
 \nonumber&= \lambda(0)\, \frac{T(Q)\, P^2 + P^2\,T(Q)}{2} + 2\ii\hbar \, \lambda^{\prime}(0)\, T(Q)\,P\\ \nonumber &+\hbar^2\left(-\lambda^{\prime\prime}(0)\, T(Q)+ \lambda^{\prime}(0)\, T^{\prime}(Q) + \frac{\lambda(0)}{4}\, T^{\prime\prime}(Q)\right)\, .
\end{align} 
We observe that the  operators \eqref{Lpm1} and \eqref{Lpm2a}  are symmetric under the condition 
\begin{equation}
\label{la0}
\lambda^{\prime}(0) = 0\, . 
\end{equation} 
Note  the appearance, in the expression of the operator \eqref{Lpm2a}, of a potential built from derivatives of the regularisation of $L(q)$. This feature is typical of quantum Hamiltonians with variable mass (see the discussion in \cite{jmll95}).

The function $\Pi(q,p)$ in~\eqref{gaussPil} can be generalized as the separable Gaussian
\begin{equation}
\label{sepgauss}
\Pi_{\sigma_{\ell} \sigma_{\wp}}(q,p) = e^{-\frac{q^2}{2\sigma_{\ell}^2}-\frac{p^2}{2\sigma_{\wp}^2}}\,,\ 
\mathfrak{F_s}\left[\Pi_{\sigma_{\ell} \sigma_{\wp}}\right](q,p)= \frac{\sigma_{\ell}\sigma_{\wp}}{\hbar}e^{-\frac{\sigma_{\wp}^2q^2}{2\hbar^2}-\frac{\sigma_{\ell}^2p^2}{2\hbar^2}}\,,
\end{equation} 
with independent widths $\sigma_{\ell}$ and $\sigma_{\wp}$, which yields to a simple formulae with familiar probabilistic content of the symplectic Fourier transform. Moreover they satisfy condition~\eqref{la0}. As was proved above, standard coherent state (or Berezin or anti-Wick)  quantization corresponds to the particular values  $\sigma_{\ell}=\sqrt{2}\ell, \sigma_{\wp}= \sqrt{2}\wp$, with the constraint  $\wp= \hbar/\ell$. On the other hand, the limit Weyl-Wigner case holds as the widths $\sigma_{\ell}$ and $\sigma_{\wp}$ are infinite. Weyl-Wigner is singular in this respect and the symplectic Fourier transform is the Dirac probability distribution centred at the origin of  the phase space. 
Recent applications of the present formalism to quantum cosmology is found in \cite{bergeronetal18,bergeronetal20}.

\section{Quantum phase portrait and its probabilistic content}
\label{probcont}

The quantization formula \eqref{quantPi1} allows to prove  the trace formula (when applicable to $f$): 
\begin{equation}
\label{traceUqp}
\mathrm{Tr}\left(U(q,p)\right)=  2\pi \hbar\delta(q,p) \, ,
\end{equation}
and so
\begin{equation}
\label{traceAf}
\mathrm{Tr}\left(A_f\right)=  \overline{\mFs}[f](0,0)=  \int_{\R^2}    f(q,p)\,  \,\frac{\ud q\,\ud p}{2\pi\hbar}\, .
\end{equation}
Thus, the operator $A_f$ is traceclass if its classical counterpart $f(q,p)$)  has finite  average on the phase space.

By using \eqref{traceAf} we derive the 
\textit{quantum phase space, i.e.  semi-classical,  portrait} of the operator as an autocorrelation averaging of the original $f$. 
 More precisely, starting from a function (or distribution) $f(q,p)$, one defines through its quantum version $A_f$ the new function $\widecheck{f}(q,p)$ as
\begin{equation}
\label{fmapcf}
\begin{split}
\widecheck{f}(q,p) &= \mathrm{Tr}\left(\mathfrak{Q}(q,p)A_f\right)\\ &=\int_{\R^2}  \, \mathrm{Tr}\left(\mathfrak{Q}(q,p)\,\mathfrak{Q}(q^{\prime},p^{\prime})\right)\, f(q^{\prime},p^{\prime})\frac{\ud q\,\ud p}{2\pi\hbar}\, .  
\end{split}
\end{equation}
The map $(q^{\prime},p^{\prime})\mapsto  \mathrm{Tr}\left(\mathfrak{Q}(q,p)\,\mathfrak{Q}(q^{\prime},p^{\prime})\right)$ might be a probability distribution if this expression is non negative.  Now, this map is better understood from the equivalent formulas,
\begin{align}
\label{fmapcf1}
\widecheck{f}(q,p)  &= \int_{\R^2}  \mFs\left[\Pi\,\widetilde\Pi\right](q^{\prime}-q,p^{\prime}-p)\, f(q^{\prime},p^{\prime}) \,\frac{\ud q^{\prime}\,\ud p^{\prime}}{2\pi\hbar} \\
\label{fmapcf2}&= \frac{1}{2\pi\hbar} \left(\overline{\mFs}\left[\Pi\,\widetilde\Pi\right]\ast f\right)(q,p) \\
\label{fmapcf3} &=\int_{\R^2}  \left(\mFs\left[\Pi\right]\ast\mFs\left[\widetilde\Pi\right]\right)(q^{\prime}-q,p^{\prime}-p)\, f(q^{\prime},p^{\prime}) \,\frac{\ud q\,\ud p}{4\pi^2\hbar^2}\, , 
\end{align}
where $\widetilde\Pi (q,p):=\Pi (-q,-p)$.
In particular, and with mild conditions on $\Pi(q,p)$,  we have for the coordinate functions, 
\begin{equation}
\label{qpqp}
 \widecheck{q}= q\, , \quad \widecheck{p}= p\,.
\end{equation}

Eq. \eqref{fmapcf1} represents the convolution ($\sim$ local averaging)  of the original $f$ with the autocorrelation of the symplectic Fourier transform of the (normalised) function $\Pi(q,p)$. 

For instance, with a separable function $\Pi(q,p)= \lambda(q)\,\mu(p)$, the quantum phase space portrait of a separable function $f(q,p) = u(q)\,v(p)$ is given by the product of two $1$d-convolutions:
\begin{equation}
\label{lmuv}
\widecheck{f}(q,p)= \left[\overline{\mathcal{F}}[\lambda \widetilde{\lambda}]\ast u\right](q)\,\left[\overline{\mathcal{F}}[\mu \widetilde{\mu}]\ast v\right](p)\,.
\end{equation}

Note that taking the dual symplectic Fourier of \eqref{fmapcf2} and applying \eqref{symfourconvqp1} yields the equality
\begin{equation}
\label{FcheckfPPFf}
\overline{\mFs}\left[\widecheck{f}\right](q,p)  = \Pi(q,p)\Pi(-q,-p)\, \overline{\mFs}\left[f\right](q,p)\,.
\end{equation}
 This alternative expression might provide the inversion formula $\widecheck{f} \mapsto f$, only if all employed formal manipulations have been mathematically justified, essentially in the framework of distributions, e.g., convolution algebra:
 \begin{equation}
\label{checkftof}
f(q,p)=\overline{\mFs}\left[ \frac{\overline{\mFs}\left[\widecheck{f}\right]}{\Pi\,\widetilde\Pi}\right](q,p)\,.
\end{equation}
For instance, suppose that we start from a  function $\widecheck{f}$ which is Lebesgue integrable on $\R^2$, i.e. $\widecheck{f}\in L^1(\R^2)$. Then  $\overline{\mFs}\left[\widecheck{f}\right]$ is continuous, bounded, and goes to $0$ at the infinity. Suppose that $1/(\Pi\,\widetilde\Pi) $ is locally integrable and  slowly increasing on $\R^2$. Then \eqref{checkftof} defines a tempered distribution on the plane. 
 
The classical limit $\widecheck{f} \to f$ of the above semi-classical formalism is also a fundamental point  to be considered. In view of the formula \eqref{fmapcf1}, it is sufficient to assert that the classical limit is reached if 
\begin{equation}
\label{classlim}
\mFs\left[\Pi\,\widetilde\Pi\right](q,p) \to 2\pi\hbar \delta(q,p)=2\pi \hbar \delta(q) \delta(p)\,, 
\end{equation}
as a certain set of introduced dimensional parameters, e.g. $\hbar$, $\ell$,  $\wp \dotsc$, go to zero. For instance, with the example of separable Gaussians 
\eqref{sepgauss}, this limit is reached at $\sigma_{\ell} \to 0$ and $\sigma_{\wp} \to 0$. 


In view of the convolution formulas \eqref{fmapcf1}-\eqref{fmapcf2}-\eqref{fmapcf3}, we are inclined to choose windows $\Pi(q,p)$, or equivalently $\mathfrak{Q}_0$,  such that
\begin{equation}
\label{probdist}
 \mFs\left[\Pi\right]
\end{equation}
is a probability distribution on the phase space $\R^2$ equipped with the measure $\dfrac{\ud q\,\ud p}{2\pi\hbar}$. 
Note that 
the uniform Weyl-Wigner choice  $\Pi(q,p)= 1$ yields 
\begin{equation}
\label{WWdist}
 \mFs\left[1\right](q,p)= 2\pi \hbar\,\delta(q,p)
\end{equation}
and $\mathfrak{Q}_0 = 2\mathrm{P}$, and  $\widecheck f= f$ in this case. Also note  that  the celebrated Wigner function  $\mathcal{W}_{\rho}(q,p)$ for a density operator or mixed quantum state $\rho$, defined by 
\begin{equation}
\label{wigA}
\mathcal{W}_{\rho}(q,p) = \mathrm{tr}\left(U(q,p)2{\sf P}U^{\dag}(q,p)\rho\right)= \mathrm{tr}\left(U(2q,2p)2{\sf P}\rho\right)\, ,
\end{equation}
is a normalised quasi-distribution which can assume negative values.

With a true probabilistic content,  the meaning of the convolution
\begin{equation}
\label{truedist}
 \mFs\left[\Pi\right]\ast\mFs\left[\widetilde\Pi\right]
\end{equation}
is clear: it is the probability distribution  for the difference of two vectors in the phase plane, viewed as independent random variables,  and thus is perfectly adapted to the abelian and homogeneous structure of the classical phase space.

We can conclude that a quantum phase space portrait in this probabilistic context  is like a measurement of  the intensity of a  diffraction pattern resulting from the function $\Pi(q,p)$ or $\mFs\left[\Pi\right](q,p)$, the coarse graining of the idealistic phase space $\R^2$. 
 
 \section{Extending Klauder's formalism with function $\Pi$}
\label{klauder}

Klauder's approach \cite{klauder12,klauder15} is based upon the comparison between the two action functionals, the classical and the quantum ones,
\begin{align}
\label{AC1}
  \texttt{A}_C  &=  \int_0^T [p(t)\dot q(t) - H_c(q(t),p(t))] \,\ud t\,,\\ 
\label{AQ2}  \texttt{A}_Q  &  = \int_0^T [\lg \psi(t)| \ii \hbar \partial/\partial t - \mathcal{H}(Q,P)]|\psi(t)\rg \,\ud t\, , 
\end{align} 
where the quantum Hamiltonian $\mathcal{H}(Q,P)$ is deduced from its  classical counterpart $H_c$ through canonical quantization, and unit norm state $|\psi\rg$ is in the domain of $\mathcal{H}(Q,P)$. Stationary variations of $\texttt{A}_C$ and $\texttt{A}_Q$  yield  their respective dynamical equations, the Hamilton ones and the Shr\"{o}dinger equation, 
\begin{align}
\label{hameq}
  &\dot q  = \{q,H_c\}= \frac{\partial H_c}{\partial p}\, , \quad    \dot p  = \{p,H_c\}= -\frac{\partial H_c}{\partial q}\, , \\
 \label{scheq}   &\ii \hbar \partial \psi/\partial t  =  \mathcal{H}(Q,P) \psi\,,
\end{align}
with solutions determined by initial conditions at $t=0$, $(q_0,p_0)$ and $|\psi_0\rg$ respectively. Hence, \eqref{scheq} results from stationary variation of 
\eqref{AQ2} over arbitrary histories of pure states $\{|\psi(t)\rg\}_0^T$ modulo fixed end points. 
The main point raised by Klauder is that ``\textit{macroscopic} observers studying a \textit{microscopic} system cannot vary histories over such a wide range''. Instead they are confined to moving the system to a new position $(q)$ or changing its velocity $(p)$ through $\dot q = \partial H_c(q,p)/\partial p$. Therefore, they are restricted to vary only the coherent  states $|q(t),p(t)\rg = U(q(t),p(t))|0\rg$, and this leads to a restricted (R) version of $\texttt{A}_Q$ given by
\begin{equation}
\label{restact}
\begin{split}
 \texttt{A}_{Q(R)}  &  = \int_0^T [\lg q(t),p(t) | \ii \hbar \partial/\partial t - \mathcal{H}(Q,P)]|q(t),p(t)\rg \,\ud t\\
                & = \int_0^T [ p(t)\dot q(t) - H(q(t),p(t))] \,\ud t\,,  \ H(q,p)= \lg q,p | \mathcal{H}(Q,P)|q,p\rg\,. 
\end{split}
\end{equation}
Here we have used the relations 
\begin{align}
\label{propdispqp1}
\hbar\dfrac{\partial}{\partial q} U(q,p) &= \left(-\ii \,P + \dfrac{\ii}{2} \,p \right) U(q,p) = -U(q,p) \left( \ii P + \dfrac{\ii}{2}\,p  \right)\, , \\
\label{propdispqp2}
\hbar\dfrac{\partial}{\partial p} U(q,p) &=  \left(\ii \, Q - \dfrac{\ii}{2} \,q \right) U(q,p) =  U(q,p) \left( \ii\,Q +  \dfrac{\ii}{2}\,q \right)\, ,
\end{align}
to prove that 
\begin{equation}
\label{pqdpq}
\ii\hbar\lg q,p |  \ud |q,p\rg = \lg 0| [(P +p )\,\ud q - Q\,\ud p|0\rg= p \,\ud q\,.
\end{equation}
We now extend this formalism in the spirit of the present work by replacing $\lg q(t),p(t) |\cdot |q(t),p(t)\rg$ in the expression \eqref{restact} with 
$\mathrm{Tr}\left(\mathfrak{Q}(q(t),p(t))\cdot\right)$ in accordance with \eqref{fmapcf}. We also replace $\mathcal{H}(Q,P)$ with $A_{H_c}$, the WH  quantized version of the classical $H_c$.   By using \eqref{propdispqp1} or \eqref{propdispqp2}, and \eqref{qpqp} we now deal with the action
\begin{equation}
\label{restactPi}
\begin{split}
 \texttt{A}^{\Pi}_{Q(R)}  &  = \int_0^T \mathrm{Tr}\left[\mathfrak{Q}(q(t),p(t))\, \left( \ii \hbar \partial/\partial t - A_{H_c}\right)\right] \,\ud t\\
                & = \int_0^T [ p(t)\dot q(t) - \widecheck{H_c}(q(t),p(t))] \,\ud t \,.  
\end{split}
\end{equation}
Stationary variations  of this  $\texttt{A}^\Pi_Q$  now yield  the semi-classical  Hamilton equations
\begin{equation}
\label{hameqPi}
  \dot q  = \{q,\widecheck{H_c}\}= \frac{\partial \widecheck{H_c}}{\partial p}\, , \quad    \dot p  = \{p,\widecheck{H_c}\}= -\frac{\partial \widecheck{H_c}}{\partial q}\, .
 \end{equation}

\section{Quantization and semi-classical portraits of truncated observables}
\label{qutrunc}

We consider  classical motions which are geometrically
 restricted to hold in some subset $E$ of the configuration space, i.e. the real line $\R$, and we 
  truncate all classical observables to $E$ as
\begin{equation}
\label{trunc}
f(q,p)\mapsto \chi_{E}(q)f(q,p)\equiv f_{\chi}(q,p)\,  ,
\end{equation}
where $\chi_{E}$ is the characteristic (or indicator) function of set $E$. Although these truncated observables are generally discontinuous, they can be quantized through  the Weyl-Heisenberg  quantization \eqref{fAf} or \eqref{quantPi1}.  We obtain the $E$-modified operators,
\begin{align} 
\label{quantchif1}
f_{\chi}(q,p)\mapsto A_{f_\chi} &= \frac{1}{2\pi \hbar}\int_{E} \ud q\int_{\R}  \ud p \,f_\chi(q,p) \mathfrak{Q}(q,p)\\
\label{quantchif2} &=  \frac{1}{2\pi \hbar}\int_{\R^2}  \ud q\,\ud p\,  \Pi(q,p)\,  \overline{\mFs}[f_\chi](q,p)\,U(q,p)\,.
\end{align}
All quantization formulae given in Section \ref{permissu} (for general $\Pi$) and in Section \ref{weight} (for separable $\Pi$) apply here with the change $f\mapsto \chi f$.
In particular,   the quantization of the singular $f_\chi(q,p)= \chi_{E}(q)$ yields   what we call the ``window'' operator, 
 \begin{equation}
\label{restId}
 A_{\chi_{E}} = \frac{1}{\sqrt{2\pi\hbar}}\int_{\R} \ud p \, \Pi(0,p)\, \widehat{\chi}(p)\,e^{\ii pQ}= \overline{\mathcal{F}}[\Pi(0,\cdot)\, \widehat{\chi}](Q)\equiv \mathcal{E}(Q)\, , 
\end{equation}
where  $\widehat{\chi}(p)=\mathcal{F}[\chi](p)$.

Note that the  Hilbert space  in which act these ``$E$-modified'' operators is left unchanged. Thus, in position representation, one continues to deal with  $\mathcal{H}= L^2(\R, \ud x)$.
Nevertheless, if the function $\Pi(q,p)$ is regular enough (resp. smooth), our approach gives rise to a regularisation (rep. smoothing) of the constraint boundary, 
i.e., a ``fuzzy'' boundary, and also a regularisation (resp. smoothing) of all discontinuous restricted observable $f_\chi(q,p)$ introduced in the quantization map \eqref{quantchif1}-\eqref{quantchif2}. 
Indeed, there is no mechanics outside the set $E\times\R$ defined by the position constraint  on the classical level. 
It is  not the same on the quantum level since our quantization method allows to go beyond the boundary of this set in a  way which can be smoothly rapidly decreasing, depending on the function $\Pi(0,p)$.

Consistently,  the semi-classical phase space portrait of the operator \eqref{quantchif1} is given by \eqref{fmapcf1}-\eqref{fmapcf2}-\eqref{fmapcf3}. Let us retain  \eqref{fmapcf2}, which appears to be the most condensed
\begin{equation}
\label{semclassA}
\widecheck{f}_\chi(q,p)  = \frac{1}{2\pi\hbar} \left(\overline{\mFs}\left[\Pi\,\widetilde\Pi\right]\ast f_\chi\right)(q,p) \, . 
\end{equation}
This function, which   is expected  to be concentrated on the classical $E\times \R$,  should be viewed as a new classical observable defined on the full phase space $\R^2$ where  $q$ and $p$ keep their status of canonical variables. 

Thus, we have here the meaningful sequence\\
 \begin{equation}
\label{regseq}
 \begin{split}
\mathrm{virtual} \ f(q,p) \rightarrow \  \mathrm{truncated} \ &f_\chi(q,p)\\
&\symbolwithin{\downarrow}{=} \\
 \mathrm{regularised} \ \widecheck{f}_\chi(q,p)  \leftarrow\  \mathrm{quantum}\  &A_{f_{\chi}}\,,
\end{split}
\end{equation}\\
allowing to establish a semi-classical dynamics \textit{\`a la} Klauder \cite{klauder12},  mainly concentrated on $E\times\R$.

\section{Operators in an interval and their semi-classical portraits}
\label{interval}
 As an elementary illustration of the formalism and as it was done with coherent states in \cite{gazkoi19}, we restrict our study to the bounded open interval $E= (a,b)$, i.e., 
\begin{eqnarray}
\chi_E(q)= \chi_{(a,b)}(q) =\Theta(q-a) - \Theta(q-b)\, ,
\label{char}
\end{eqnarray}
where $\Theta(q)=  \chi_{(0,+\infty)}(q)$ is the Heaviside function. 

\begin{figure}[H]
\centering
\includegraphics[width=0.3\textwidth]{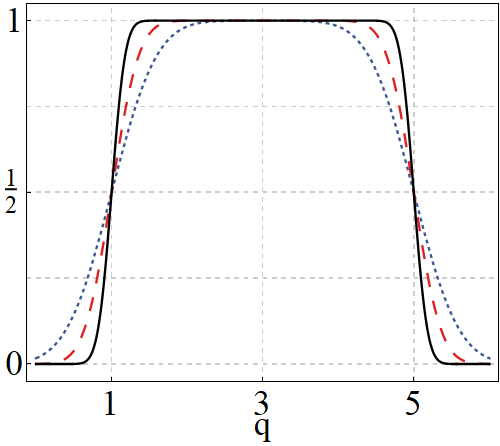}
\caption{\footnotesize{(In units of $\hbar=1$) Semi-classical portraits $\widecheck{\chi}^{(G)}_{(a,b)}(q)$ of the characteristic function fixed at $a=1$ and $b=5$, together with $\sigma_{\wp}=3,5,10$ (blue-dotted, red-dashed and black-solid, respectively).}}
\label{FchiG}
\end{figure}

At this point, we should be aware that the motion in our bounded geometry is determined by a confinement potential, such as an oscillator-like interaction, and the presence of position-dependent mass terms. Before considering the details of the classical model under consideration, it is worth to discuss first the general properties of the  functions $\Pi(q,p)$ used throughout the rest of the text. To this end, let us consider a function $\Pi_{G}(q,p)$, together with the respective symplectic Fourier transform $\mFs[\Pi_{G}\,\widetilde{\Pi}_{G}](q,p)$, of the form
\begin{equation}
\begin{split}
\Pi_{G}(q,p)&=e^{-\frac{q^{2}}{2\sigma_{\ell}^{2}}}e^{-\frac{p^{2}}{2\sigma_{\wp}^{2}}}e^{\frac{\gamma}{2}qp} \, , \\ \mFs[\Pi_{G}\,\widetilde{\Pi}_{G}](q,p)&=\frac{\sigma_{\ell}\sigma_{p}}{2\Lambda\hbar}e^{-\frac{\Lambda^{2}_{\ell}q^{2}}{4\hbar^2}}e^{-\frac{\Lambda_{\wp}^{2}p^{2}}{4\hbar^2}}e^{\frac{\Lambda_{0}}{\hbar^{2}}qp} \, ,
\end{split}
\label{nsepG}
\end{equation}
where the parameters in $\mFs[\Pi_{G}\,\widetilde{\Pi}_{G}](q,p)$ are given by
\begin{align}
\label{nsepG1}
&\Lambda_{\ell}^{2}=\frac{\sigma_{\wp}^{2}}{\Lambda^{2}}, \quad \Lambda_{\wp}^{2}=\frac{\sigma_{\ell}^{2}}{\Lambda^{2}} , \quad \Lambda_{0}=\frac{\sigma^{2}_{\ell}\sigma_{\wp}^{2}\gamma}{4\Lambda^{2}}, \\
\label{gamma}
&\Lambda^{2}=\frac{4-\sigma_{\ell}^{2}\sigma_{\wp}^{2}\gamma^{2}}{4}>0,
\end{align}
with $\sigma_{\ell}$ and $\sigma_{\wp}$ positive constants with dimension of length and momentum, respectively, and the real coupling constant $\gamma$ with inverse action dimension constrained to $\vert\gamma\vert<2/(\sigma_{\ell}\sigma_{\wp})$. Notice that~\eqref{nsepG} is a generalization of the separable function~\eqref{sepgauss}, we thus call $\Pi_{G}(q,p)$ as the \textit{non-separable Gaussian} function. In general, the quantum operator $\hat{\chi}^{(G)}_{(a,b)}$ related to the characteristic function and the semi-classical portrait $\widecheck{\chi}_{(a,b)}^{(G)}(q)$ are respectively defined as
\begin{equation}
\hat{\chi}^{(G)}_{(a,b)}(Q)\equiv B_{\frac{\sigma_{\wp}}{\sqrt{2}}}(a,b;Q), \quad \widecheck{\chi}^{(G)}_{(a,b)}(q)\equiv B_{\frac{\sigma_{\wp}}{2}}(a,b;q),
\label{chiG}
\end{equation}
with
\begin{equation}
B_{\sigma_\wp}(a,b;x):= \frac{1}{2}\left( \operatorname{Erfc}\left[ \frac{\sigma_{\wp}}{\hbar}
(x-b)\right] - \operatorname{Erfc}\left[ \frac{\sigma_{\wp}}{\hbar}
(x-a)\right] \right) \, ,
\label{Bsig}
\end{equation}
where Erfc$(z)$ stands for the \textit{complementary error function}~\cite{abraste72,Olv10}. Notice that both expressions in~\eqref{chiG} are independent of the coupling factor $\gamma$, that is, the non-separability of $\Pi_{G}(q,p)$ plays no role in the quantization of the characteristic function. Nevertheless, as we discuss in the sequel, the effects of $\gamma$ are clear once we consider more general classical observables. Interestingly, the regularization introduced by $\Pi(q,p)$ leads to a smooth function $\widecheck{\chi}^{(G)}_{(a,b)}(q)$ that approximates the original characteristic function~\eqref{char}. The latter allows defining the appropriate quantum operator $\hat{\chi}^{(G)}_{(a,b)}(Q)$, and consequently any truncated observable. The behavior of the characteristic function is depicted in Fig.~\ref{FchiG} for several values of $\sigma_{\wp}/\hbar$, where it is clear that $\widecheck{\chi}^{(G)}_{(a,b)}(q)$ converges to~\eqref{char} at $\sigma_{\wp}/\hbar\to\infty$.

\section{Some applications: Position-dependent mass models in a bounded interval}
\label{potentials}

A well-known problem in the quantization procedure arises in determining the quantum operators related to classical observables of the form $f(q,p)=q^{n}p^{m}$. Given that the canonical position and momentum operators do not commute~\eqref{ccr}, the operator ordering of the resulting quantum model is not unique~\cite{Bor25,Bor27,Spr83,Cah69a,Cah69b,Gos16}. A prime example is given by the quantization of classical models associated with position-dependent mass (PDM) terms~\cite{jmll92,jmll95}, where different quantization rules lead to different quantum Hamiltonians with different spectrum, see~\cite{Cre88} for details. Despite the ordering problem, PDM models find interesting application in the study of heterostructures~\cite{jmll74,Gor69,Roo83}, and recently in the study of supersymmetric models in quantum mechanics~\cite{quesne09,cruzycruz08,cruzrosas09,cruzrosas11}.

\begin{figure}
\centering
\subfloat[][]{\includegraphics[width=0.3\textwidth]{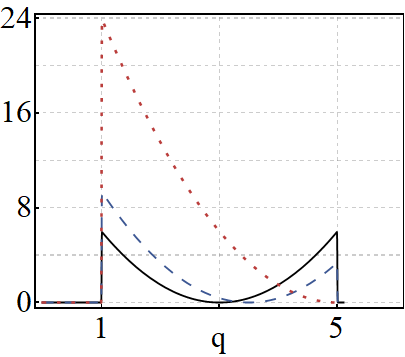}
\label{VchiC}}
\hspace{5mm}
\subfloat[][]{\includegraphics[width=0.3\textwidth]{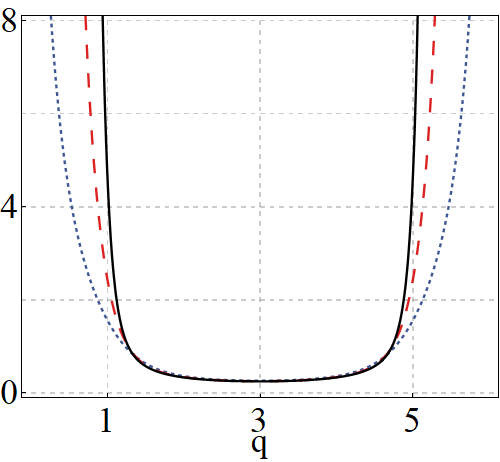}
\label{masschiG}}
\caption{\footnotesize{(a) Classical truncated-potential $V_{\chi}(q)$ given in~\eqref{truncHc1} for $V_{0}=3$, together with $q_{0}=3$ (solid-black), $q_{0}=3.5$ (dashed-blue), and $q_{0}=5$ (dotted-red). (b) (In units of $\hbar=1$) Regularized semi-classical mass term $m^{(G)}_{\chi}(q)$, given in~\eqref{QmGchi}, for $\sigma_{\wp}=3,5,10$ (dotted-blue, dashed-red and solid-black, respectively).}}
\end{figure}

Throughout this section, to illustrate the usefulness of the WH covariant integral quantization, we consider a classical PDM model corresponding to the family of nonlinear oscillators of the form
\begin{equation}
H(q,p)=\frac{p^{2}}{2m(q)}+\frac{V_{0}}{2}(q-q_{0})^{2}, \quad m(q)=\frac{m_{0}L^{2}}{(q-a)(b-q)}, \quad m_{0}>0.
\label{classpdm}
\end{equation}
The latter arises as a modification of the nonlinear oscillator discussed in~\cite{mathew74}. The constant $L$ is a characteristic length, e.g. $L=b-a$. In this form, the related quantum operators are determined with the appropriate use of the $\Pi(q,p)$ function. Although, from~\eqref{classpdm}, we notice that $m(q)$ is a physical mass-term only in the interval $q\in(a,b)$, that is, $m(q)$ is a positive function. To preserve the physical structure of the model~\eqref{classpdm}, we introduce the truncated classical Hamiltonian of the form
\begin{equation}
H_{\chi}(q,p)\equiv\chi_{(a,b)}(q) H(q,p)=\frac{p^{2}}{2m_{\chi}(q)}+V_{\chi}(q), 
\label{truncHc}
\end{equation}
where the truncated position-dependent mass and potential terms are respectively given by
\begin{equation}
\label{truncHc1}
\begin{split}
&\frac{1}{m_{\chi}(q)}\equiv\mathfrak{m}_{\chi}(q)=\frac{(q-a)(b-q)}{m_{0}L^{2}}\chi_{(a,b)}(q), \\
&V_{\chi}(q)=\frac{V_{0}}{2}(q-q_{0})^{2}\chi_{(a,b)}(q).
\end{split}
\end{equation}
From the constraint in the bounded interval, we realize that the shape of the truncated oscillator potential $V_{\chi}(q)$ is steered by the potential minimum $q_{0}$. For details, see Fig.~\ref{VchiC}.

Before proceeding, it is useful to compute the quantum operator and semi-classical portrait related to the mass-term given in~\eqref{truncHc1}. Following~\eqref{quantchif2} and~\eqref{semclassA}, we notice that the function
\begin{equation}
\label{mGchi}
\begin{split}
\mathfrak{M}_{\sigma_{\wp}}(x)&=\frac{1}{m_{0}L^2}\left[(x-a)(b-x)-\frac{\hbar^2}{\sigma_{\wp}^{2}} \right]B_{\frac{\sigma_{\wp}}{\sqrt{2}}}(a,b;x) \\
&+\frac{1}{m_{0}L^2}\frac{\hbar}{\sqrt{2\pi\sigma_{\wp}^{2}}}\left[ (x-a)e^{-\frac{\sigma_{\wp}^{2}}{2\hbar^2}(x-b)^{2}}-(x-b)e^{-\frac{\sigma_{\wp}^{2}}{2\hbar^2}(x-a)^{2}} \right],
\end{split}
\end{equation}
leads to the respective quantum and semi-classical mass terms conveniently written in the form
\begin{equation}
\frac{1}{M^{(G)}_{\chi}(Q)}\equiv\widehat{\mathfrak{m}}_{\chi}(Q)=\mathfrak{M}_{\sigma_{\wp}}(Q), \quad \frac{1}{m^{(G)}_{\chi}(q)}\equiv\widecheck{\mathfrak{m}}_{\chi}(q)=\mathfrak{M}_{\frac{\sigma_{\wp}}{\sqrt{2}}}(q)\,.
\label{QmGchi}
\end{equation}
The behavior of $m^{(G)}_{\chi}(q)$ is depicted in Fig.~\ref{masschiG} for several values of the Gaussian width $\frac{\sigma_{\wp}}{\hbar}$. Notice that, due the regularization effects of $\Pi_{G}(q,p)$, the singularities of the initial mass-term $m(q)$ in~\eqref{classpdm} have been smoothed in the semi-classical portrait, leading to a function that extends beyond the bounded interval $q\in(a,b)$. In turn, the classical limit is recovered for $\sigma_{\wp}/\hbar\rightarrow\infty$, where the function $B_{\sigma}(a,b;q)$ converges, in the sense of a distribution, to the characteristic function~\eqref{char}.

For the non-separable Gaussian function $\Pi_{G}(q,p)$, it can be shown that a truncated classical function $f_{\chi}(q,p)=p^{2}h(q)\chi_{(a,b)}(q)$ gives rise to the semi-classical portrait $\widecheck{f}_{p^{2}h_{\chi}}$ of the form
\begin{equation}
\widecheck{f}_{p^{2}h_{\chi}(q)}=\widecheck{f}_{h_{\chi}}\left(p+\hbar^{2}\gamma\frac{\widecheck{f}_{h_{\chi}}'}{\widecheck{f}_{h_{\chi}}} \right)^{2}+\widecheck{f}_{h_{\chi}}\left[\hbar^{4}\gamma^{2}\left(\frac{\widecheck{f}_{h_{\chi}}'}{\widecheck{f}_{h_{\chi}}}\right)'+ \frac{2\hbar^{2}}{\sigma_{\ell}^{2}} \right]
\label{semclassGph}
\end{equation}
with $\widecheck{f}_{h_{\chi}}\equiv\widecheck{f}_{h_{\chi}}(q)$ the semi-classical portrait of $f(q,p)=h(q)\chi_{(a,b)}(q)$, and the notation $\widecheck{f}'=\frac{\partial \widecheck{f}}{\partial q}$ denotes the partial derivative of any semi-classical function $\widecheck{f}$ with respect to $q$. 

From~\eqref{semclassGph}, together with $1/m^{(G)}_{\chi}$ in~\eqref{QmGchi}, we can easily compute the semi-classical portrait of the truncated Hamiltonian~\eqref{classpdm}. After some calculations we get
\begin{equation}
\widecheck{H}^{(G)}_{\chi}(q,p):=\frac{1}{2m^{(G)}_{\chi}}\left(p-\hbar^{2}\gamma\frac{[m_{\chi}^{(G)}]'}{m_{\chi}^{(G)}} \right)^{2}+\widecheck{V}_{eff}^{(G)}(q),
\label{semclassGHtrun}
\end{equation}
with $[m^{(G)}_{\chi}]'\equiv \partial m^{(G)}_{\chi}/\partial q$, and $\widecheck{V}_{eff}^{(G)}(q)$ an effective potential term of the form
\begin{multline}
\widecheck{V}_{eff}^{(G)}(q):=\frac{V_{0}}{2}\left[(q-q_{0})^{2}+\frac{2\hbar^{2}}{\sigma_{\wp}^{2}}\right]\widecheck{\chi}^{(G)}_{(a,b)}(q)\\ 
+\frac{1}{2m^{(G)}_{\chi}}\left[-\hbar^{4}\gamma^{2}\left(\frac{[m^{(G)}_{\chi}]'}{m^{(G)}_{\chi}}\right)'+\frac{2\hbar^{2}}{\sigma_{\ell}^{2}}\right]\\
-\frac{V_{0}\hbar}{\sqrt{4\pi\sigma_{\wp}^{2}}}\left[(q+b-2q_{0})e^{-\frac{\sigma_{\wp}^{2}}{4\hbar^{2}}(q-b)^{2}}-(q+a-2q_{0})e^{-\frac{\sigma_{\wp}^{2}}{4\hbar^{2}}(q-a)^{2}}\right].
\label{effpotG}
\end{multline}
The effective potential is composed of several parts. Besides the effects of the truncated oscillator-like interaction, the effective potential contains the effects of the geometrical constraint imposed on the classical model, which is reflected in the regularized characteristic function $\widecheck{\chi}_{(a,b)}(q)$. Also, the initial mass $m(q)$ contributes with an additional regularized term in the potential through $m^{(G)}_{\chi}(q)$, and its derivatives. Finally, the effects of the coupling factor $\gamma$ are present in the potential as well, which also induces a \textit{minimal coupling} term~\cite{Gol11} in the kinetic energy term of the semi-classical Hamiltonian~\eqref{semclassGHtrun}. This is an interesting phenomenon that can be interpreted as the existence of an effective vector potential of the form $\vec{A}(q)=\hbar^{2}\gamma[m^{(G)}_{\chi}]'/m^{(G)}_{\chi} \hat{n}$, where $\hat{n}$ is a unit vector in the same direction as the canonical coordinate $q\hat{n}$. Since we are dealing with a one-dimensional system, the direction of the vector potential is not relevant, and henceforth will be omitted. 

We want to stress out that a striking feature of the semi-classical portrait relies upon the Hamiltonian structure, that is, we can determine the dynamics of the semi-classical model through the Hamilton equations of motion~\eqref{hameqPi}. Although the equations of motion would lead to a nonlinear differential equation associated with the position $q(t)$, as it happens with the classical PDM model of~\cite{mathew74}, we may still get information about the dynamics of the system from its respective phase-space trajectories. The latter is achieved by fixing the total energy of the system as a constant $\widecheck{H}_{\chi}^{(G)}(q,p)=E$. Such a procedure is a legitimate operation, since the semi-classical Hamiltonian is time-independent, and thus the energy is a conserved quantity~\cite{cruzrosas13,Gol11}. It is worth notice that, from the phase-space trajectories themselves, we can not obtain information about the direction of motion of the particle. We thus compute the vector quantity $\vec{\mathcal{F}}$ defined in terms of the canonical variables as~\cite{Gol11}
\begin{equation}
\vec{\mathcal{F}}=(\dot{q}(q,p),\dot{p}(q,p)),
\label{field}
\end{equation}
where $\dot{q}(q,p)$ and $\dot{p}(q,p)$ are determined from the Hamilton equations of motion~\eqref{hameq}. We have 
\begin{equation}
\dot{q}(q,p)=\frac{1}{m^{(G)}_{\chi}}\left(p-\hbar^{2}\gamma\frac{[m^{(G)}_{\chi}]'}{m^{(G)}_{\chi}}\right),
\label{Gqdot}
\end{equation}
\begin{equation}
\begin{split}
\dot{p}=\frac{[m^{(G)}_{\chi}]'}{2[m^{(G)}_{\chi}]^2}&\left[ p-\hbar^{2}\gamma\left(3\frac{[m^{(G)}_{\chi}]'}{\widecheck{m}_{\chi}}-2\frac{[m^{(G)}_{\chi}]''}{[m^{(G)}_{\chi}]'}\right)\right] \\
&\times\left(p-\hbar^{2}\gamma\frac{[m^{(G)}_{\chi}]'}{m^{(G)}_{\chi}}\right)
-\frac{\partial\widecheck{V}_{eff}^{(G)}}{\partial q}.
\label{Gforce}
\end{split}
\end{equation}
In this form, by combining the phase-space trajectories for a fixed total energy, together with the vector field $\vec{\mathcal{F}}$, we obtain a complete description of the system dynamics, we discuss such details in the sequel. 

Alternatively, from~\eqref{Gqdot}, the canonical momentum $p$ is determined in terms of the position-dependent mass term and the particle velocity. This allows us to obtain an equivalent representation of the system dynamics through the reparametrized Hamiltonian
\begin{equation}
\widecheck{\mathfrak{h}}_{\chi}^{(G)}(q,\dot{q})\equiv \widecheck{H}^{(G)}_{\chi}(q,p(q,\dot{q}))=\frac{m^{(G)}_{\chi}\dot{q}^{2}}{2}+\widecheck{V}_{eff}^{(G)}(q) \, .
\label{Hqdot}
\end{equation}

From the general results obtained so far, it is worth separating the discussion in some particular cases.
\subsection{Separable case $\gamma=0$}
\label{sepWg0}
\begin{figure}
\subfloat[][]{\includegraphics[width=0.3\textwidth]{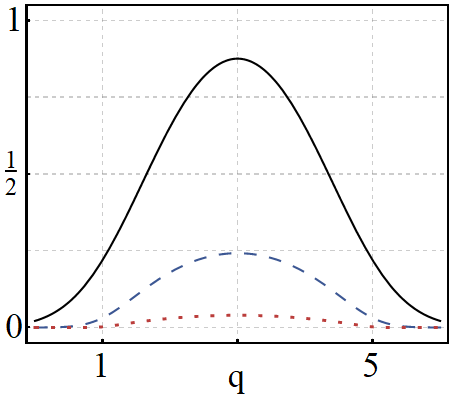}
\label{VchiG0}}
\hspace{5mm}
\subfloat[][]{\includegraphics[width=0.3\textwidth]{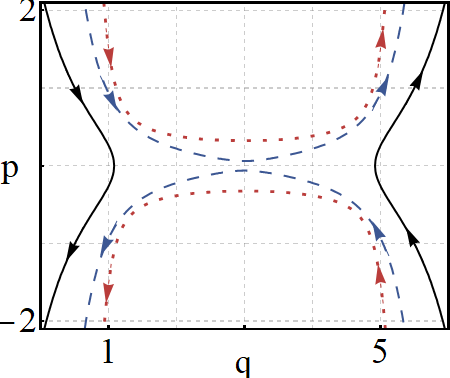}
\label{phaseG0}}
\caption{\footnotesize{(a) Effective potential $V_{eff}^{(G)}(q)$ with a null oscillator-like interaction $V_{0}=0$, the parameters are fixed to $a=1$, $b=5$, $m_{0}L^{2}=1$, $\gamma=0$. (b) Phase-space trajectories of the potentials depicted in~(a) for a total energy $\widecheck{H}^{(G)}_{\chi}=E=1/4$, the arrows depict $\vec{\mathcal{F}}$ given in~\eqref{field}. In all the cases, we have used units of $\hbar=1$ with the Gaussian width fixed at $\sigma_{\wp}=\sigma_{\ell}=2$ (solid-black), $\sigma_{\wp}=\sigma_{\ell}=4$ (dashed-blue) and $\sigma_{\wp}=\sigma_{\ell}=6$ (dotted-red).}}
\end{figure}
As we have previously pointed out, for $\gamma=0$, the function $\Pi_{G}(q,p)$ reduces to the separable one introduced in~\eqref{sepgauss}. Additionally, let us also consider a null oscillator-like interaction $V_{0}=0$. The latter leads, in general, to a non-null effective potential that arises from the bounded interval of the initial model, as in the case discussed in~\cite{gazkoi19}, but in our model we also have the effects of the regularized PDM term $1/m^{(G)}_{\chi}$. The behavior of the effective potential is depicted in Fig.~\ref{VchiG0}, from where it is evident that $\widecheck{V}_{\chi}^{(G)}$ is regularized beyond the initial bounded interval $q\in(a,b)$. Given the lack of returning points of the potential, the particle is not allowed to perform a bounded motion, this fact is confirmed by the information obtained from the phase-space trajectories (in units of $\hbar=1$), with a fixed total energy $\widecheck{H}_{\chi}^{(G)}(q,p)=E=1/4$, depicted in Fig.~\ref{phaseG0}. For $\sigma_{\wp}=2$, the particle can not pass from one region of the interval to the other, since the potential energy prohibits such a transition. Moreover, the particle tends to be attracted to either of the regularized walls generated by the mass-term. At the same energy $E=1/4$, but for higher values of $\sigma_{\wp}$, the particle has enough energy to transit from one wall to the other, nevertheless, the particle never returns to the initial wall, that is, the system does not allow bouncing behavior. Notice that in the classical limit, $\sigma_{\wp}/\hbar\rightarrow\infty$ and $\sigma_{\ell}/\hbar\rightarrow\infty$, the effective potential vanishes inside the bounded interval.
\begin{figure}
\centering
\hspace{1mm}
\subfloat[][]{\includegraphics[width=0.35\textwidth]{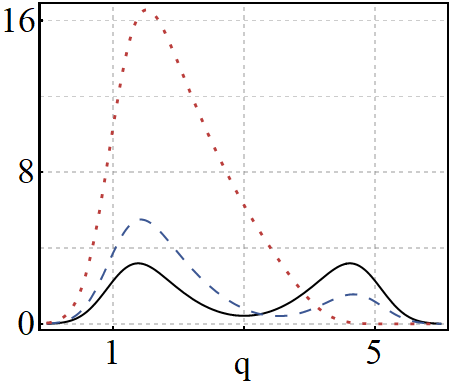}
\label{VchiG1}}
\hspace{5mm}
\subfloat[][]{\includegraphics[width=0.35\textwidth]{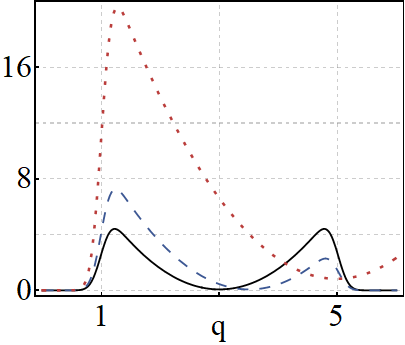}
\label{VchiG2}}
\caption{\footnotesize{(In units of $\hbar=1$) Regularized semi-classical effective potential $V_{eff}^{(G)}(q)$ with the regularization parameters $\sigma_{\ell}=\sigma_{\wp}=4$ (a) and $\sigma_{\ell}=\sigma_{\wp}=10$ (b). The rest of parameters are $V_{0}=3$, $a=1$, $b=5$, $m_{0}=1$. The position of the potential minimum takes the values $q_{0}=3$ (solid-black), $q_{0}=3.5$ (dashed-blue), and $q_{0}=5$ (dotted-red).}}
\label{potG}
\end{figure}

In turn, for $V_{0}>0$, we have different scenarios. For instance, if $a<q_{0}<b$, the oscillator-like interaction gets truncated in such a way that classical confinement is possible for finite range of energies $E_{m}<E<E_{th}$, with $E_{m}$ and $E_{th}$ the minimum and threshold energies, respectively. The behavior of the effective potential is depicted in Fig.~\ref{potG} for several values of the potential minimum position $q_{0}$ and the regularization parameters $\sigma_{\ell}$ and $\sigma_{wp}$. In particular, in Fig.~\ref{VchiG1}, for $q_{0}=3$ (solid-black) we can see the effects of the regularized truncated potential, where a symmetric potential is obtained, and energies of classical confinement are present. For $q_{0}=3.5$ (dashed-blue), the effective potential becomes asymmetric, and the energy threshold $E_{th}$ is lower compared with the symmetric case. For increasing ratios $\sigma_{\wp}/\hbar$ and $\sigma_{\ell}/\hbar$ the effective potential displays a behavior more similar to the classical one, as shown in Fig.~\ref{VchiG2}. Clearly, in the classical limit, we recover the initial truncated-potential previously shown in Fig.~\ref{VchiC}.

\begin{figure}
\centering
\subfloat[][]{\includegraphics[width=0.3\textwidth]{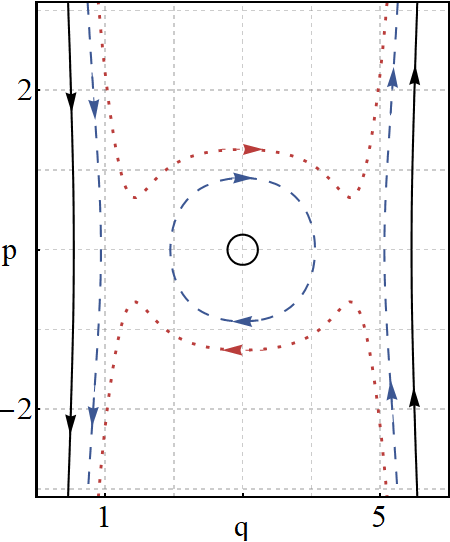}
\label{phaseG1}}
\hspace{1mm}
\subfloat[][]{\includegraphics[width=0.3\textwidth]{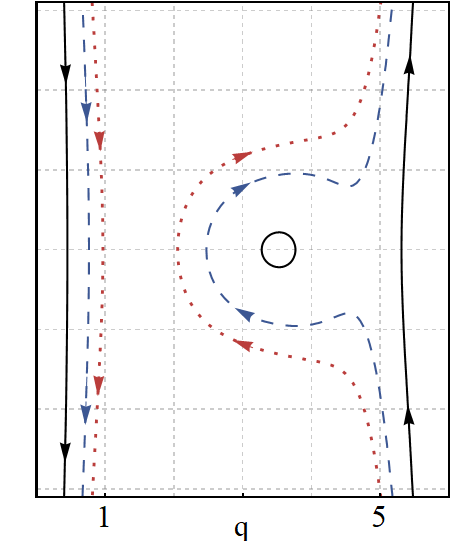}
\label{phaseG2}}
\hspace{1mm}
\subfloat[][]{\includegraphics[width=0.3\textwidth]{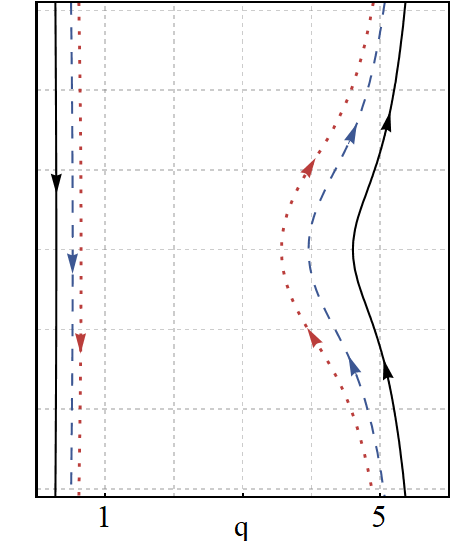}
\label{phaseG3}}
\caption{\footnotesize{(In units of $\hbar=1$) Phase-space trajectories of $\widecheck{H}_{\chi}^{(G)}(q,p)\equiv E$, given in~\eqref{semclassGHtrun}, for the separable case $\gamma=0$ and the parameters $a=1$, $b=5$, $\sigma_{\ell}=\sigma_{\wp}=4$, $V_{0}=3$, $m_{0}L^{2}=1$. The curves correspond to energies $E=0.5$ (solid-black), $E=2$ (dashed-blue), and $E=3.5$ (dotted-red). The arrows depict $\vec{\mathcal{F}}$ given in~\eqref{field}. The position minimum is being fixed as $q_{0}=3$ (a), $q_{0}=3.5$ (b) and $q_{0}=5$.}}
\label{phaseGsep}
\end{figure}

The classical confinement in the potential is interpreted in the phase-space description as closed trajectories. The latter is depicted in Figs.~\ref{phaseG1}-\ref{phaseG2} for several different energies. Contrary to an oscillator interaction, in our case the bounded motion depends on the initial condition as well. Additionally, comparing the dashed-blue trajectories for $q_{0}=3$ (Fig.~\ref{phaseG1}) and that for $q_{0}=3.5$ (Fig.~\ref{phaseG2}), with both being fixed at the same energy $E=2$, it is clear that the presence of classical confinement strongly depends on the position of the potential minimum $q_{0}$. Moreover, in the critical cases $q_{0}=a$ and $q_{0}=b$, the effective potential prohibits the existence of confinement, as depicted in Fig.~\ref{phaseG3}.

\begin{figure}
\centering
\subfloat[][]{\includegraphics[width=0.3\textwidth]{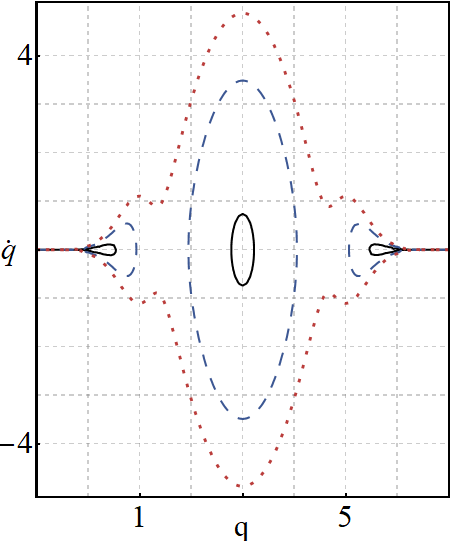}
\label{phaseqqdot1}}
\hspace{1mm}
\subfloat[][]{\includegraphics[width=0.3\textwidth]{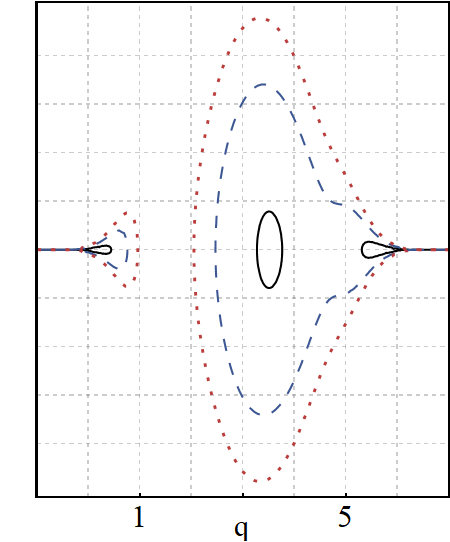}
\label{phaseqqdot2}}
\hspace{1mm}
\subfloat[][]{\includegraphics[width=0.3\textwidth]{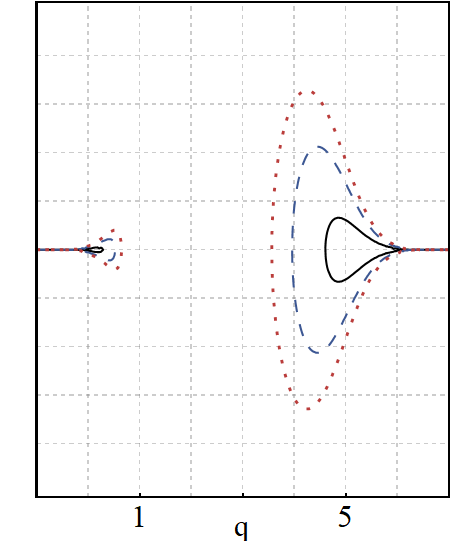}
\label{phaseqqdot3}}
\caption{\footnotesize{(In units of $\hbar=1$) Trajectories $\widecheck{\mathfrak{h}}_{\chi}^{(G)}(q,\dot{q})=E$, given in~\eqref{Hqdot}, for the separable case $\gamma=0$ and the parameters $a=1$, $b=5$, $\sigma_{\ell}=\sigma_{\wp}=4$, $V_{0}=3$, $m_{0}L^{2}=1$. The curves correspond to energies $E=0.5$ (solid-black), $E=2$ (dashed-blue), and $E=3.5$ (dotted-red). The position minimum is being fixed as $q_{0}=3$ (a), $q_{0}=3.5$ (b) and $q_{0}=5$.}}
\label{phaseQdot1}
\end{figure}

Alternatively, we can analyze the dynamics of the particle when it approaches the regularized walls. To this end, we depict in Fig.~\ref{phaseQdot1} the trajectories of $\widecheck{\mathfrak{h}}_{\chi}^{(G)}$ in the  $q$-$\dot{q}$ plane, computed from~\eqref{Hqdot}. In the latter, for the sake of clarity, we have used the same parameters as in Fig.~\ref{phaseGsep}. As mentioned above, the particle never returns from the walls, such a conclusion is confirmed from Figs.~\ref{phaseqqdot1}-\ref{phaseqqdot1}, where the velocity $\dot{q}$ drops close to zero as the particle approaches either of the walls. Also, the particle is allowed to travel beyond the initial bounded interval $[a,b]$, because of the regularization effects of $\Pi_{G}(q,p)$, where it spends an infinite amount of time, that is, the particle never returns. Additional details are provided in the sequel.

\subsection{Non-separable case $\gamma\neq 0$}
\label{nsepGg}
\begin{figure}
\centering
\subfloat[][$q_{0}=3$]{\includegraphics[width=0.3\textwidth]{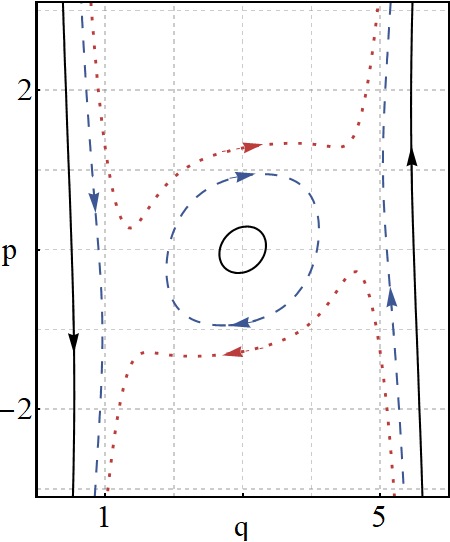}
\label{phaseG4}}
\hspace{1mm}
\subfloat[][$q_{0}=3.5$]{\includegraphics[width=0.3\textwidth]{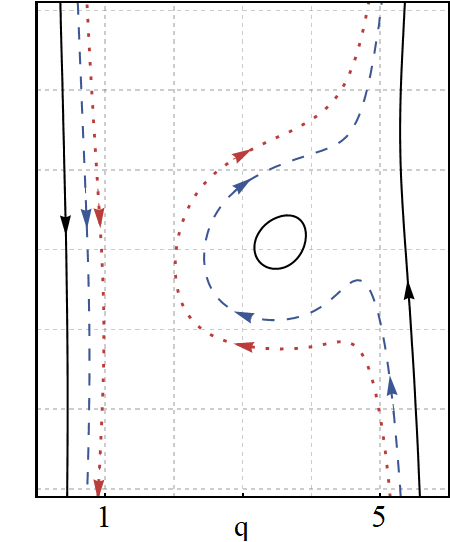}
\label{phaseG5}}
\hspace{1mm}
\subfloat[][$q_{0}=5$]{\includegraphics[width=0.3\textwidth]{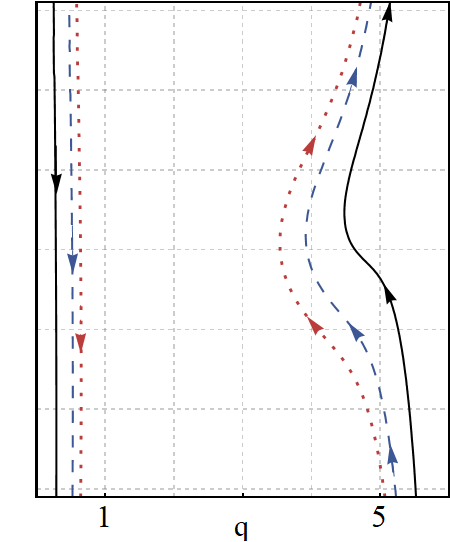}
\label{phaseG6}}
\caption{\footnotesize{(In units of $\hbar=1$) Phase-space trajectories of $\widecheck{H}_{\chi}^{(G)}(q,p)=E$, given in~\eqref{semclassGHtrun}, for the non-separable case $\gamma=0.3$ and the parameters $a=1$, $b=5$, $\sigma_{\ell}=\sigma_{\wp}=4$, $V_{0}=3$, $m_{0}L^{2}=1$. The curves correspond to energies $E=0.5$ (solid-black), $E=2$ (dashed-blue), and $E=3.5$ (dotted-red). The arrows depict $\vec{\mathcal{F}}$ given in~\eqref{field}.}}
\label{phaseGnsep}
\end{figure}

In this case, the global behavior of the effective potential is similar to that of the previous case depicted in Fig.~\ref{potG}. Nevertheless, the most relevant change comes from the kinetic energy term of the Hamiltonian~\eqref{semclassGHtrun} in the form of a minimal coupling. To get more insight in the dynamics of the system under the influence of such a minimal coupling, we depict the phase-space trajectories in Fig.~\ref{phaseGnsep}, with the same parameters as in Fig.~\ref{phaseGsep}. It is worth to recall that the coupling factor is bounded by the inequality~\eqref{gamma}. Given that the Gaussian widths have been fixed (in units of $\hbar=1$) as $\sigma_{\wp}=4$ and $\sigma_{\ell}=4$, we thus choose the coupling factor as $\gamma=0.1<1/8$. With the latter, we can see the differences that arises from the non-separability of the function $\Pi_{G}(q,p)$. From Figs.~\ref{phaseG4}-\ref{phaseG6}, it is clear that the trajectories, both open and closed, are deviated with respect to the $\gamma=0$ case. As we have pointed out previously, the minimal coupling is related to a vector potential of the form $\vec{A}(q)=\gamma[m^{(G)}_{\chi}]'/m^{(G)}_{\chi} \hat{n}$, for which there is an associated magnetic field $\vec{B}=\nabla\times\vec{A}$. It is worth to recall that the minimal coupling term is a purely quantum effect, for it is proportional to $\hbar$ that vanishes in the classical limit.

\begin{figure}
\centering
\subfloat[][]{\includegraphics[width=0.3\textwidth]{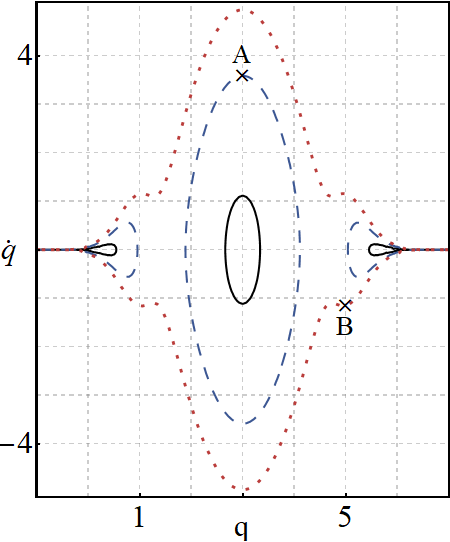}
\label{phaseAqqdot1}}
\hspace{1mm}
\subfloat[][]{\includegraphics[width=0.3\textwidth]{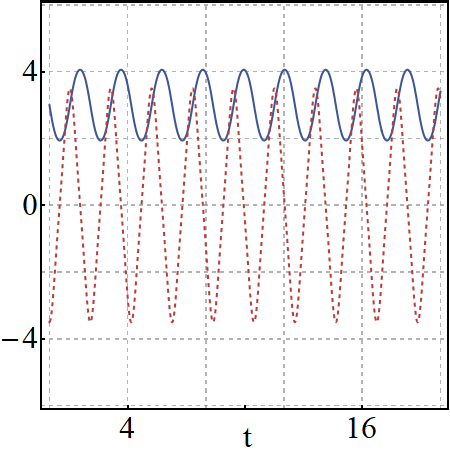}
\label{phaseAqqdot2}}
\hspace{1mm}
\subfloat[][]{\includegraphics[width=0.3\textwidth]{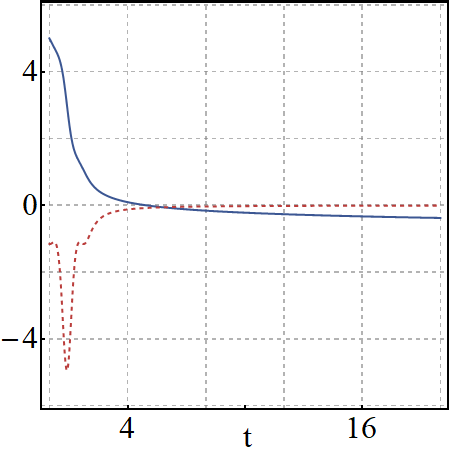}
\label{phaseAqqdot3}}
\caption{\footnotesize{(In units of $\hbar=1$) (a) Trajectories of $\widecheck{\mathfrak{h}}_{\chi}^{(G)}(q,\dot{q})=E$ , determined from~\eqref{Hqdot}, for the non-separable function $\Pi(q,p)$ with the parameters $a=1$, $b=5$, $V_{0}=3$, $m_{0}=L^{2}=1$, together with $\sigma_{\ell}=\sigma_{\wp}=4$ and $\gamma=0.1$. The curves correspond to energies $E=0.5$ (solid-black), $E=2$ (dashed-blue), and $E=3.5$ (dotted-red). (b-c) Numerical solutions for the position $q(t)$ (solid-blue) and velocity $\dot{q}(t)$ (dashed-red) for the initial conditions $A$ and $B$ depicted in (a)}}
\label{phaseQdot2}
\end{figure}

We complement this section by discussing the trajectories in the $q$-$\dot{q}$ plane, depicted in Fig.~\ref{phaseAqqdot1}. From $\widecheck{\mathfrak{h}}^{(G)}_{\chi}(q,\dot{q})$ given in ~\eqref{Hqdot} we see that the minimal coupling term $A(q)$ does not contribute in the dynamics, but the effects of the non-separability due the coupling factor $\gamma$ are still present in the effective potential $\widecheck{V}_{eff}(q)$. To further understand the dynamics, we can compute from~\eqref{hameq} the equation of motion for $q(t)$ with ease, but the resulting equation is nonlinear, and determining an analytic solution for $q(t)$ is not possible. Nevertheless, we can overcome such an issue by performing some numerical calculations. Such information is presented in Fig.~\ref{phaseAqqdot2}-\ref{phaseAqqdot3} for the initial conditions $A$ and $B$ depicted in Figs.~\ref{phaseAqqdot2}-\ref{phaseAqqdot3}, respectively. From those figures we see that, in Fig.~\ref{phaseAqqdot2}, the initial conditions allow a closed trajectory, whereas in Fig.~\ref{phaseAqqdot3} the particle spends an infinite amount of time traveling through the wall.

\subsection{Quantum regularized PDM model}
\label{GQmodel}
For completeness, we determine the quantum model related to the truncated Hamiltonian~\eqref{truncHc}. We proceed similarly as in the semi-classical approach, and after some calculations we get 
\begin{equation}
\widehat{H}_{\chi}^{(G)}:=\frac{1}{2} \left\{\frac{1}{2M^{(G)}_{\chi}(Q)},\left( P-\frac{\hbar^{2}\gamma}{2}\frac{[M^{(G)}_{\chi}(Q)]'}{M^{(G)}_{\chi}(Q)}\right)^{2} \right\}+ \hat{V}_{eff}^{(G)}(Q),
\label{QHchiG}
\end{equation}
where $\{\hat{A},\hat{B}\}=\hat{A}\hat{B}+\hat{B}\hat{A}$ stands for the anti-commutator, and we have used the notation
\begin{equation}
[M^{(G)}_{\chi}]'\equiv \left. \frac{\partial}{\partial x} \frac{1}{\mathfrak{M}_{\sigma_{\wp}}(x)}\right\vert_{x\rightarrow Q} \, ,
\end{equation}
with $\mathfrak{m}_{\sigma_{\wp}}(x)$ given in~\eqref{mGchi}. The term $\hat{V}_{eff}^{(G)}(Q)$ stands for a quantum effective potential term of the form
\begin{multline}
\hat{V}^{(G)}_{eff}(Q):=\frac{V_{0}}{2}\left[(Q-q_{0})^{2}+\frac{\hbar^{2}}{\sigma_{\wp}^{2}} \right]\hat{\chi}_{(a,b)}^{(G)}(Q)\\+\frac{1}{M^{(G)}_{\chi}(Q)}\left[-\frac{\hbar^{4}\gamma^{2}}{4}\left( \frac{[M^{(G)}_{\chi}(Q)]'}{M^{(G)}_{\chi}(Q)} \right)' +\frac{\hbar^{2}}{\sigma_{\ell}^{2}}\right]\\
-\frac{V_{0}\hbar}{\sqrt{2\pi\sigma_{\wp}^{2}}}\left[(Q+b-2q_{0})e^{-\frac{\sigma_{\wp}^{2}}{2\hbar^{2}}(Q-b)^{2}}-(Q+a-2q_{0})e^{-\frac{\sigma_{\wp}^{2}}{2\hbar^{2}}(Q-a)^{2}}\right].
\end{multline}
Notice that, the kinetic energy term in~\eqref{QHchiG} corresponds to a symmetrical PDM ordering introduced by Gora-Williams~\cite{Gor69} in the study of band-structures in slowly graded semiconductors, and later generalized by von-Roos~\cite{Roo83}. Nevertheless, our model possesses an additional minimal coupling term that depends on the regularized quantized mass-term $M^{(G)}_{\chi}$.

As discussed in Section~\ref{CIQWH}, the apodization $\Pi(q,p)$ is equivalent to a trace-class density operator $\mathfrak{Q}_{0}$, where both quantities are related by~\eqref{IWHtr}. From our model, the non-separable Gaussian function~\eqref{sepgauss} allows determining a unique density operator $\mathfrak{Q}^{(G)}_{0}$. In the Fock basis $\{\vert n \rangle\}_{n=0}^{\infty}$, the upper-diagonal matrix elements $\mathfrak{Q}^{(G)}_{0;(n+2M,n)}\equiv\langle n+2M\vert\mathfrak{Q}^{(G)}_{0}\vert n\rangle$, with $n,M=0,1,\cdots$, are given in units of $\hbar=1$ by
\begin{equation}
\begin{aligned}
\mathfrak{Q}^{(G)}_{0;(n+2M,n)}&=\frac{\sqrt{n!(n+2M)!}(2M)!}{2^{2M}\Delta_{\ell}\Delta_{\wp}\Delta_{\wp}^{2M}} \left( \mathfrak{G}_{1}-\frac{2i\gamma}{\Delta_{\ell}^{2}}\mathfrak{G}_{2} \right) \, ,
\end{aligned}
\label{matQ0G}
\end{equation}
where
\begin{equation}
\nonumber
\begin{aligned}
\mathfrak{G}_{1}:=\sum_{k=0}^{M}&\sum_{r=0}^{n}\sum_{s=0}^{r}\frac{(-1)^{r+k}\left(\tfrac{1}{2}+k\right)_{s}\left(\tfrac{1}{2}+N-k\right)_{r+s}}{s!k!(r-s)!(M-k)!(n-r)!(2M+r)!}\left(\frac{\Delta_{\wp}}{\Delta_{\ell}}\right)^{2k+2s}\\
&\times\frac{1}{\Delta_{\wp}^{2r}}\,{}_{2}F_{1}\left(\left.\begin{split} \tfrac{1}{2}+s+k,\tfrac{1}{2}+M+r-s-k \\ \tfrac{1}{2} \end{split}\right\vert \frac{\gamma^{2}}{4\Delta_{\ell}^{2}\Delta_{\wp}^{2}} \right)
\end{aligned}
\end{equation}
\begin{equation}
\nonumber
\begin{aligned}
\mathfrak{G}_{2}:=\sum_{k=0}^{M}&\sum_{r=0}^{n}\sum_{s=0}^{r}\frac{(-1)^{r+k}\left(\tfrac{3}{2}+k\right)_{s}\left(\tfrac{1}{2}+N-k\right)_{r+s}}{s!k!(r-s)!(M-k-1)!(n-r)!(2M+r)!}\left(\frac{\Delta_{\wp}}{\Delta_{\ell}}\right)^{2k+2s}\\
&\times\frac{1}{\Delta_{\wp}^{2r}}\,{}_{2}F_{1}\left(\left.\begin{split} \tfrac{3}{2}+s+k,\tfrac{1}{2}+M+r-s-k \\ \tfrac{3}{2} \end{split}\right\vert \frac{\gamma^{2}}{4\Delta_{\ell}^{2}\Delta_{\wp}^{2}} \right)
\end{aligned}
\end{equation}
with $\Delta^{2}_{\ell}=\tfrac{1}{2}+\tfrac{1}{\sigma_{\ell}^{2}}$, $\Delta^{2}_{\wp}=\tfrac{1}{2}+\tfrac{1}{\sigma_{\wp}^{2}}$. The self-adjointness of $\mathfrak{Q}_{0}$ implies that the lower-diagonal elements are $\mathfrak{Q}^{(G)}_{0;(n,n+2M)}:=\left[\mathfrak{Q}^{(G)}_{0;(n+2M,n)}\right]^{*}$. In particular, after fixing the parameters to
\begin{equation}
\sigma^{2}_{\ell}=2\ell^{2}\frac{\kappa_{R}}{\vert \kappa\vert^{2}} \, ,\quad \sigma^{2}_{\wp}=2\wp^{2}\kappa_{R} \, , \quad \gamma=-\frac{\kappa_{I}}{\kappa_{R}} \, ,
\label{slsp-squeezed}
\end{equation}
with $\kappa$ given in~\eqref{sqgaussPil}, we recover the density operator related to the squeezed vacuum state, that is, $\mathfrak{Q}_{0}^{(\xi)}=S(\xi)\vert \psi_{\ell} \rangle\langle \psi_{\ell} \vert S^{\dagger}(\xi)$, with $S(\xi)$ defined in~\eqref{sqvacuum}. We still have freedom in the remaining parameters in~\eqref{slsp-squeezed}, from which an interesting case is achieved by considering $\kappa=1$ ($\xi=\eta=0$) and $\sigma_{\ell}=\sigma_{\wp}=\sigma>0$, this leads to the diagonal density operator
\begin{equation}
\mathfrak{Q}_{0}^{th}=\frac{1}{\Delta^{2}}\sum_{n=0}^{\infty}\left(1-\frac{1}{\Delta^{2}}\right)^{n}\vert n \rangle\langle n\vert \, , \quad \Delta^{2}=\frac{1}{2}+\frac{1}{\sigma^{2}} \, ,
\label{Q0Gther}
\end{equation}
which is nothing but a thermal state, as discussed in~\cite{bergaz14}. Finally, the most simple case is determined after fixing $\sigma=\sqrt{2}$ ($\Delta=1$), such that $\mathfrak{Q}^{0}_{0}=\vert \psi_{\ell} \rangle\langle \psi_{\ell} \vert$. In this form, we recover the coherent state quantization previously discussed in the literature~\cite{bergaz14}. 

\section{Conclusion}
\label{conclu}
In this work, we have explored a quantization mechanism for classical systems defined in a finite interval based on the Weyl-Heisenberg covariant integral quantization. The essential feature of this approach lies in the use of a smooth function $\Pi(q,p)$, defined in the whole classical phase-space $\mathbb{R}^{2}$, that regularizes the discontinuities present in the classical model. As a result, we obtain quantum models defined in terms of genuinely self-adjoint operators in the respective Hilbert space, and therefore there is no need to include self-adjoint extensions in terms of unbounded operators. Instead, we have at our disposal a set of regularization parameters such that the quantum model can be adjusted to the physical system under consideration while preserving the regularity of the operators. The latter is an essential feature of our approach. On the other hand, the $\Pi(q,p)$ function defines a unique density operator $\mathfrak{Q}$ such that, after averaging the operators, leads to a semi-classical portrait of the quantum model, that is, a classical-like model in which $\hbar$ is still present. In this form, the formalism of Klauder~\cite{klauder12,klauder15} is extended such that the action, defined in terms of the semi-classical portrait of the quantum Hamiltonian, is minimized. From the latter, a set of Hamilton equations is determined, which allows us to treat the semi-classical system within the well-known Hamiltonian formalism.

An interesting application of the quantization procedure is provided by a classical model with an oscillator interaction and a position-dependent mass term whose physicality (positive-definitive mass) is only valid in a finite interval. In this form, the truncation of the domain to a constrained geometry allows defining the resulting classical model properly, from which we obtain effective infinite-walls at the end of the interval domain. In such a case, a non-separable Gaussian function $\Pi(q,p)$ regularizes the discontinuities in the truncated potential and the mass-term. For instance, in the semi-classical level, an oscillatory motion is obtained for total mechanical energies lower than the effective potential threshold. The latter depends strongly on the regularization parameters and the domain of the finite-interval. It is explicitly shown that, for some intervals, oscillatory motion is completely forbidden, since the energy threshold is zero. The particle travels toward either of the walls, after which it never returns. The maximal trapping is achieved for finite-intervals such that the effective potential becomes symmetric with respect to its minimum, that is, the energy threshold reaches its maximum value. In those cases, for mechanical energies below the threshold, the particle performs harmonic motion. In contrast, for mechanical energies higher than the threshold, the particle escapes the confinement potential and travels toward one of the walls.

A striking feature of the non-separable function $\Pi(q,p)$ is given by the appearance of a minimal coupling in the form of an effective magnetic potential term for both the regularized quantum and semi-classical Hamiltonians. This is a purely quantum effect that arises from the non-separability of $\Pi(q,p)$, which is controlled by a coupling parameter. In this form, the resulting quantum model is governed by a generalization of the PDM Hamiltonian with the Gora-Williams ordering~\cite{Gor69}. Therefore, the minimal coupling vanishes in either the classical limit or by tuning the coupling the parameter in such a way that $\Pi(q,p)$ becomes a separable function. In this regard, it is worth exploring in detail which other properties would arise from the non-separability of the function $\Pi(q,p)$. Results in this direction will be reported elsewhere.

\appendix

\section{Harmonic analysis on $\C$ or $\R^2$ and symbol calculus}
\label{fourier}
We give here all formulas with $\hbar=1$.
\subsection{In terms of $z=\frac{q+\ii p}{\sqrt{2}}$ and $\bar z=\frac{q-\ii p}{\sqrt{2}}$}
\begin{itemize}
  \item Symplectic Fourier transform on $\mathbb{C}$ (for a sake of simplicity, we write $f(z,\bar z) \equiv f(z)$)
  \begin{align}
 \label{symFourTr1}  \mathfrak{f_s}[f](z)&=\int_{\mathbb{C}} e^{ z \bar \xi -\bar z \xi} f(\xi)\,  \frac{\ud^2 \xi}{\pi}= \int_{\mathbb{C}} e^{2\ii \,\mathrm{Im}(z \bar \xi)} f(\xi)\,  \frac{\ud^2 \xi}{\pi}\\
\label{symFourTr2}   &=\int_{\mathbb{C}} e^{z\circ\xi} f(\xi)\,  \frac{\ud^2 \xi}{\pi}\quad \mbox{(notation often used here)}\, . 
     \end{align}
\item  Dirac-Fourier formula
\begin{equation}
\label{diracfourier}
\mathfrak{f_s}[1](z)= \int_{\mathbb{C}}e^{z \circ \xi }\, \dfrac{\ud^2 \xi}{\pi} = \int_{\R^2}e^{-\ii (qy-px)}\, \dfrac{\ud x\,\ud y}{2\pi} =  2\pi \delta(q)\,\delta(p)=\pi \delta^{2}(z)\, .
\end{equation}
\item The symplectic Fourier transform is its inverse: it is an involution
\begin{equation}
\label{ff1}
\mathfrak{f_s}[\mathfrak{f_s}[f]](z) = f(z) \  \Leftrightarrow \ \mathfrak{f_s}\,\mathfrak{f_s} = \mathfrak{f_s}^2 = I\, .
\end{equation}
\item The symplectic Fourier transform commutes with the parity operator
\begin{equation}
\label{invsymFourTr2}
\mfs= \sfP \,\mfs \,\sfP \, , \quad (\sfP\,f)(z)=f(-z)= \tilde{f}(z)\,, \ \tilde{f}(z):= f(-z)\, . 
\end{equation}
\item Reflected symplectic Fourier transform
\begin{equation}
\label{refsymFourTr1} \overline{\mathfrak{f_s}}[f](z)=\int_{\mathbb{C}} e^{-z\circ\xi} f(\xi)\,  \frac{\ud^2 \xi}{\pi}=  \mathfrak{f_s}[f](-z)=  \mathfrak{f_s}\left[\tilde{f}\right](z)= \overline{\mathfrak{f_s}\left[\bar{f}\right](z)}\, . 
\end{equation}
\item The reflected symplectic Fourier transform is its inverse
\begin{equation}
\label{rfrf1}
\omfs \,\omfs = I \,. 
\end{equation}
\item Factorization of the parity operator
\begin{equation}
\label{factP}
\omfs\mfs = \mfs \omfs = \sfP\,. 
\end{equation}

\item Symplectic Fourier transform and translation\\
with 
\begin{equation}
\label{translation}
(\mathsf{t}_z\,f)(z^{\prime}):= f(z^{\prime}-z)\, , 
\end{equation}
\begin{equation}
\label{symfourtransl}
\mfs\left[\mathsf{t}_z\,f\right](z^{\prime})= e^{z^{\prime}\circ z}\, \mfs[f](z^{\prime})\,,\quad \omfs\left[\mathsf{t}_{-z}\,f\right](z^{\prime})= e^{z^{\prime}\circ z}\, \omfs[f](z^{\prime}) \,. 
\end{equation}
\item Symplectic Fourier transform and derivation
\begin{align}
\label{symFourder1}
   \frac{\partial^k}{\partial z^k} \,\mfs[f](z)&= \mfs\left[\bar\xi^k\, f\right](z)\, ,  \quad  &\frac{\partial^k}{\partial z^k} \,\omfs[f](z)= \omfs\left[\left(-\bar\xi\right)^k\, f\right](z) \, ,     \\
 \label{symFourder2}    \frac{\partial^k}{\partial \bar z^k} \,\mfs[f](z)&= \mfs\left[(-\xi)^k\,f\right](z)\, , \quad &\frac{\partial^k}{\partial \bar z^k} \,\omfs[f](z)= \omfs\left[\xi^k\,f\right](z)\, ,   \\
 \label{symFourder3}  \mfs\left[\frac{\partial^k}{\partial \xi^k} \,f\right](z)&= \bar z^k\,\mfs\left[f\right](z)\, , \quad &\omfs\left[\frac{\partial^k}{\partial \xi^k} \,f\right](z)=(- \bar z)^k\,\omfs\left[f\right](z)\, ,  \\
 \label{symFourder4}   \mfs\left[\frac{\partial^k}{\partial \bar\xi^k} \,f\right](z)&= (-z)^k\,\mfs\left[f\right](z)\, , \quad &\omfs\left[\frac{\partial^k}{\partial \bar\xi^k} \,f\right](z)=z^k\,\omfs\left[f\right](z)\, . 
\end{align}
\item Convolution product with complex variables
\begin{equation}
\label{convolC}
(f\ast g)(z):= \int_{\C}\ud^2z^{\prime}\, f(z-z^{\prime}) \, g(z^{\prime})= (g\ast f)(z)\, . 
\end{equation}
\item Symplectic Fourier transform of convolution products
\begin{align}
\label{symfourconv1}
  \mfs[f\ast g] (z)&= \pi \,  \mfs[f] (z)\, \mfs[ g] (z)\, ,  \\
 \label{symfourconv2}    \mfs[f\, g] (z)&= \frac{1}{\pi} \,  (\mfs[f] \ast \mfs[ g]) (z)\, . 
\end{align}
\item Symplectic Fourier transform of Gaussian
\begin{equation}
\label{symfourgauss}
\mfs\left[e^{\nu \, \vert\xi\vert^2}\right](z) = \frac{1}{(-\nu)}\, e^{ \frac{\vert z\vert^2}{\nu}}= \omfs\left[e^{\nu \, \vert\xi\vert^2}\right](z)\, , \quad \mathrm{Re}(\nu) < 0\, . 
\end{equation}
 \item Symplectic Fourier transform of  $D$  
  \begin{equation}
\label{ftransfD}
\int_{\C} e^{z \circ z^{\prime}} \, D(z^{\prime})\,\frac{\ud^2 z^{\prime}}{\pi}  = 2\, D(2z)\,{\sf P} = 2 {\sf P}\, D(-2z)  \,.
\end{equation}
\end{itemize}
\subsection{In terms of $q$ and $p$}
\begin{itemize}
  \item In terms of coordinates $z= (q+\ii p)/\sqrt{2}$, $\xi= (x+\ii y)/\sqrt{2}$,
\begin{equation}
\label{symFourqp1}
\mathfrak{f_s}[f](z)\equiv \mathfrak{F_s}[F](q,p)= \int_{\R^2}e^{-\ii (qy - px)}\, F(x,y)\,\frac{\ud x\,\ud y}{2\pi} = \mathfrak{F}[F](-p,q)\, , 
\end{equation}
where $ \mathfrak{F}$ denotes the standard two-dimensional Fourier transform,
\begin{equation}
\label{stFourqp}
\mathfrak{F}[F](k_x,k_y)= \int_{\R^2}e^{-\ii (k_x x + k_y y)}\, F(x,y)\,\frac{\ud x\,\ud y}{2\pi}\, , 
\end{equation}
with inverse
\begin{equation}
\label{stFourqpinv}
\overline{\mathfrak{F}}[F](k_x,k_y)= \int_{\R^2}e^{\ii (k_x x + k_y y)}\, F(x,y)\,\frac{\ud x\,\ud y}{2\pi}= \mathfrak{F}[F](-k_x,-k_y)\, .
\end{equation} 
  \item $\mathfrak{F_s}$ is involutive, $\mathfrak{F_s}\left[\mathfrak{F_s}[F]\right]= \mathfrak{F_s}^2[F]= F$ like its ``dual'' defined as
  \begin{equation}
\label{dsymFourqp}
\overline{\mathfrak{F_s}}[F](q,p)= \mathfrak{F_s}[F](-q,-p)=\int_{\R^2}e^{\ii (qy - px)}\, F(x,y)\,\frac{\ud x\,\ud y}{2\pi} = \mathfrak{F}[F](p,-q)\, , 
\end{equation}
 \item Symplectic Fourier transform and derivation
\begin{align}
\label{symFourderqp1}
   \frac{\partial^k}{\partial q^k} \,\mFs[F](q,p)&= (-\ii)^k \mFs\left[y^k\, F\right](q,p)\, ,    &\frac{\partial^k}{\partial q^k} \,\omFs[F](q,p)= \ii^k\mFs\left[y^k\, F\right](q,p) \, ,     \\
 \label{symFourderqp2}   \frac{\partial^k}{\partial p^k} \,\mFs[F](q,p)&= \ii^k\mFs\left[x^k\, F\right](q,p)\, ,  &\frac{\partial^k}{\partial p^k} \,\omFs[F](q,p)= (-\ii)^k\omFs\left[x^k\,F\right](q,p)\, ,   \\
 \label{symFourderqp3}  \ii^k\mFs\left[\frac{\partial^k}{\partial x^k} \,F\right](q,p)&= p^k\,\mfs\left[F\right](q,p)\, ,  &(-\ii)^k\omFs\left[\frac{\partial^k}{\partial x^k} \,f\right](q,p)=p^k\,\omFs\left[F\right](q,p)\, ,  \\
 \label{symFourderqp4}  (-\ii)^k \mFs\left[\frac{\partial^k}{\partial y^k} \,F\right](q,p)&= q^k\,\mFs\left[F\right](q,p)\, , &\ii^k\omFs\left[\frac{\partial^k}{\partial y^k} \,F\right](q,p)=q^k\,\omFs\left[F\right](q,p)\, . 
\end{align}
\item Convolution product 
\begin{equation}
\label{convolC1}
(F\ast G)(q,p):= \int_{\C}\ud q^{\prime}\,\ud p^{\prime}\, F(q-q^{\prime}, p-p^{\prime}) \, G(q-q^{\prime},p-p^{\prime})= (G\ast F)(z)\, . 
\end{equation}
\item Symplectic Fourier transform of convolution products
\begin{align}
\label{symfourconvqp1}
  \mFs[F\ast G] (q,p)&= 2\pi \,  \mFs[F] (q,p)\, \mFs[ G] (q,p)\, ,  \\
 \label{symfourconvqp2}    \mFs[F\, G] (q,p)&= \frac{1}{2\pi} \,  (\mFs[F] \ast \mFs[ G]) (q,p)\, . 
\end{align}
\item Same formulae for $\omFs$
\end{itemize}

\subsection*{Acknowledgments}
J.-P. Gazeau acknowledges partial support of CNRS-CRM-UMI 3457. V. Hussin acknowledges the support of research grants from NSERC of Canada. K. Zelaya acknowledges the support from the Mathematical Physics Laboratory of the Centre de Recherches Mat\'ematiques, through a postdoctoral fellowship. He also acknowledges the support of
Consejo Nacional de Ciencia y Tecnolog\'ia (Mexico), grant number A1-S-24569.


\begin{thebibliography}{99}
\bibitem{apodi} An apodization function (also called a tapering function or window function) is a function used to smoothly bring a sampled signal down to zero at the edges of the sampled region.
\end{thebibliography}
\end{document}